\documentclass[a4paper,fleqn]{cas-dc}

\def\tsc#1{\csdef{#1}{\textsc{\lowercase{#1}}\xspace}}
\tsc{WGM}
\tsc{QE}
\tsc{EP}
\tsc{PMS}
\tsc{BEC}
\tsc{DE}

\usepackage[authoryear]{natbib}
\usepackage{amsmath,amssymb}
\usepackage{graphicx}
\usepackage{subfigure} 
\usepackage{xcolor} 
\usepackage{hyperref}
\usepackage{tcolorbox}
\usepackage{algorithm,algpseudocode,float}
\usepackage{savesym}
\savesymbol{checkmark}
\savesymbol{carriagereturn}
\usepackage{dingbat}
\usepackage{mathtools}

\newcommand\simiid{\stackrel{\mathclap{\normalfont\mbox{\tiny \tiny i.i.d}}}{\sim}}

\newcommand\independent{\protect\mathpalette{\protect\independenT}{\perp}}
\def\independenT#1#2{\mathrel{\rlap{$#1#2$}\mkern2mu{#1#2}}}

\newcommand\rurl[1]{%
  \href{http://#1}{\nolinkurl{#1}}%
}

\begin{document}

\let\WriteBookmarks\relax
\def\floatpagepagefraction{1}
\def\textpagefraction{.001}
\shorttitle{Title...
}
\shortauthors{Parametric Surface Projection Method for Derivatives Portfolio Risk Management }

\title[mode = title]{Temporal Volatility Surface Projection
\\[-1pt] {\Large Parametric Surface Projection Method for Derivatives Portfolio Risk Management}
\\[20pt]
}

\author[1]{ \textcolor{black}{Shiva Zamani}}

\author[2]{ \textcolor{black}{Alireza Moslemi Haghighi}}

\author[3]{ \textcolor{black}{Hamid Arian}}

\address[1]{Sharif University of Technology, Teymoori Sq, Tehran, Iran 1459973941}
\address[3]{York University, 4700 Keele St, Toronto, Ontario, Canada M3J 1P3}



\begin{abstract}
This study delves into the intricate realm of risk evaluation within the domain of specific financial derivatives, notably options. Unlike other financial instruments, like bonds, options are susceptible to broader risks. A distinctive trait characterizing this category of instruments is their non-linear price behavior relative to their pricing parameters. Consequently, evaluating the risk of these securities is notably more intricate when juxtaposed with analogous scenarios involving fixed-income instruments, such as debt securities. A paramount facet in options risk assessment is the inherent uncertainty stemming from first-order fluctuations in the underlying asset's volatility. The dynamic patterns of volatility fluctuations manifest striking resemblances to the interest rate risk associated with zero-coupon bonds. However, it is imperative to bestow heightened attention on this risk category due to its dependence on a more extensive array of variables and the temporal variability inherent in these variables. This study scrutinizes the methodological approach to risk assessment by leveraging the implied volatility surface as a foundational component, thereby diverging from the reliance on a singular estimate of the underlying asset's volatility.
\end{abstract}


\maketitle

\section{Introduction}\label{chap:intro}

Within the financial industry, identifying and mitigating risks emerge as paramount concerns. The inherent uncertainty in the market engenders multifaceted manifestations of risk. Notably, liquidity, market, and credit risks are particularly consequential (\cite{hull2012risk}). Each risk category considers discrete origins of uncertainty, thereby necessitating the formulation of corresponding models. This study focuses on the sphere of market risk concerning the option market.

Market risk\footnote{Throughout this study, we employ Value-at-Risk (VaR) as the primary risk measure for market risk assessment. However, it is essential to note that the methods presented in this research allow for the estimation of the empirical distribution of portfolio value changes, enabling the calculation of both VaR and Expected Shortfall (ES).}, stemming from the uncertainty of market price fluctuations, is a widely acknowledged concern in equities and fixed-income securities. However, regarding derivatives, risk assessment becomes notably more complex. This study is dedicated to risk analysis, specifically emphasizing European options associated with non-dividend-paying stocks. Two primary approaches are utilized to estimate market risk effectively in the context of options: Historical Simulation (HS) and Parametric Estimation.

Historical simulation is a straightforward method for estimating VaR in option portfolios, using real market data to capture non-normal return distributions. It facilitates scenario analysis (\cite{hull2012risk}) to understand portfolio performance under various market conditions. However, its accuracy depends on data quality and doesn't consider forward-looking information or market dynamics changes. It assumes past statistical properties persist into the future, and it may be challenging to apply to illiquid assets or limited trading history (\cite{pritsker2006hidden}). Additionally, it's constrained by historical data length. Combining it with other VaR methods can mitigate these limitations, which we have also done.

The parametric approach in risk estimation focuses primarily on the underlying pricing model and endeavors to predict and estimate VaR by leveraging structural information. Initially, this approach entails calibrating the chosen model, providing valuable insights into the model's parameters. Subsequently, following the calibration procedure, the calculation of VaR is executed through conditional simulation (\cite{glasserman2000efficient}). As such, the parametric approach often employs the Monte Carlo (MC) method in conjunction with it. The estimation of VaR is rooted in the empirical distribution of returns, which is adjusted or conditioned based on the outcomes derived from the calibration process. This conditioning step aims to account for potential alterations in market dynamics, volatility levels, or correlations, which might not be adequately captured through the simplistic historical simulation approach.

The Black-Scholes model within the framework of parametric estimation is the cornerstone of our research. The introduction of the pricing model by \cite{black1973pricing} marked a seminal advancement in trading European option contracts. This specific model relies on several simplifying assumptions akin to various models. Notably, the constant volatility in the underlying asset constitutes one such assumption, a facet that scholars endeavored to incorporate into the broader Black-Scholes framework in subsequent years. 

In the context of portfolio analysis, the absence of uncertainty related to volatility prompted the development of VaR models. Examples of such models include the Delta-Gamma method introduced by \cite{jorion2007value} and the straightforward one-day Monte Carlo method. In both approaches, volatility is considered constant over a single period, resulting in negligible volatility-driven risk within that timeframe. However, these calculations may lack precision due to the simplifying assumption of constant volatility inherent in the Black-Scholes model. 

Historical data frequently reveals that asset price volatility is not consistent but rather fluctuates between high and low volatility periods. \cite{mandelbrot1967distribution} initially introduced this concept in their pioneering research. Hence, while the Black-Scholes model provides reasonable estimates for many options, it may produce inaccurate results when applied to options on assets with highly variable or stochastic volatility, such as during financial crises or significant news events. Additionally, historical events, including the notorious Black Monday of 1987, prompted market participants to reevaluate their reliance on this model (\cite{hull2003options}). These events engendered a realization that the model might need to be more aptly suited to account for the considerable fluctuations and volatility witnessed in the underlying asset. Consequently, the applicability of the Black-Scholes model comes into question in situations marked by significant market turbulence, potentially leading traders utilizing this framework to face considerable financial losses. 
\begin{figure}
    \centerline{\includegraphics[width=\linewidth]{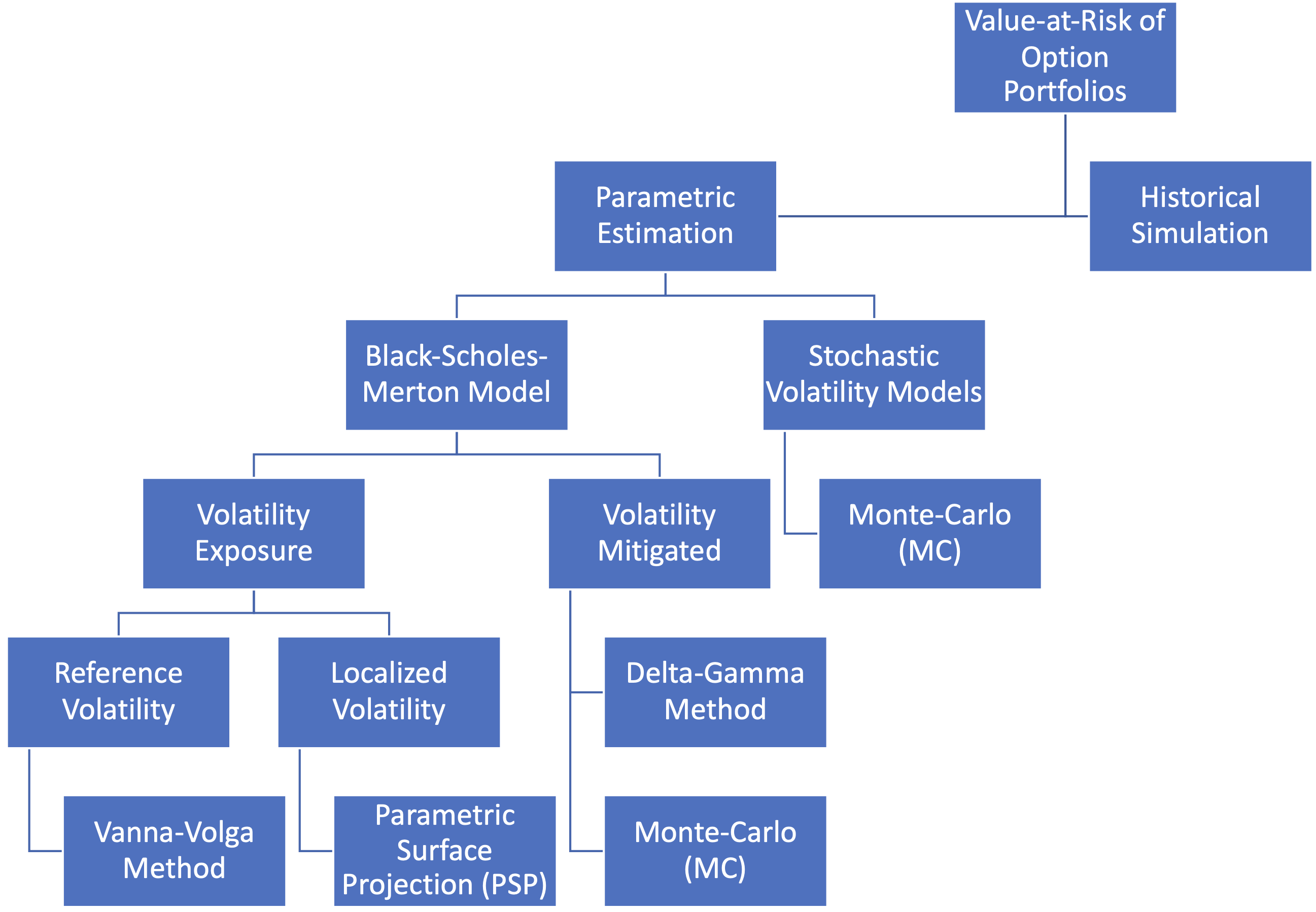}}
    \caption{This figure illustrates the proposed hierarchy of VaR calculation methods within option portfolios, as introduced in the introduction. It is important to note that the hierarchy does not imply any temporal precedence of one method over another. Instead, it visually represents how our suggested model relates to other methods in this context.}
    \label{fig:Intro_Models}
\end{figure}

In response to the limitations associated with the constant volatility assumption, options pricing has developed advanced models to refine methodologies. Notable contributions in this regard include \cite{hull1987pricing} and \cite{dupire1997pricing} pioneering researches, introducing stochastic volatility and local volatility models, respectively. These models represent significant strides toward reconciling the disparities arising from constant volatility assumptions when juxtaposed with the intricate dynamics of real-world markets.

However, the introduction of such advanced models has brought forth new challenges. Calibration processes for these models can be highly intricate, demanding a nuanced approach. Additionally, accurately forecasting the temporal fluctuations in volatility remains a formidable task, as highlighted in \cite{Verma2008ImpliedVM}. Furthermore, there are concerns regarding the compatibility of these advanced models with the second fundamental law of asset pricing, as discussed by \cite{clark2011foreign}.

Despite these challenges, these models offer promising avenues for addressing the volatility-related uncertainties intrinsic to options pricing. Moreover, in estimating VaR, these models often involve conditional simulation methods within the Monte Carlo framework, enhancing their applicability in capturing the dynamic nature of financial markets.

Researchers and market participants often employ implied volatility within the Black-Scholes framework to harness the simplicity of the Black-Scholes model and address the uncertainties stemming from volatility changes. Implied volatility serves as a means to gauge anticipated future volatility in the underlying asset, thus constituting a pivotal component of the Black-Scholes model. This estimation draws from the market prices of options, prompting market participants to base option quotations on their implied volatility rather than relying solely on prevailing market pricing (\cite{hull2003options}).

\cite{carr2020option} influential work delineates two distinct approaches for assessing option risks within the Black-Scholes framework: the reference volatility model and localized volatility models. 

The reference volatility model is designed to establish a consistent pricing framework applicable to a broad spectrum of options by introducing the concept of reference volatility. For example, in their study, \cite{mercuriovanna} employed the vanna-volga model to delve into Black-Scholes pricing, subsequently facilitating VaR calculations. This model incorporates the idea of reference volatility, offering a systematic approach to pricing various options.

On the contrary, localized volatility models pivot on the implied volatility of individual options, aiming to deduce pricing implications without a deep understanding of the underlying rationale. Within this context, \cite{carr2020option} introduced the top-down valuation approach, while \cite{arslan2009gamma} and \cite{gershon2018model} extended the vanna-volga framework. These methodologies collectively integrate the implied volatility of individual options into the valuation process. This enables a more nuanced assessment of option pricing dynamics, providing insights into market sentiment and potential mispricings.

Consistent with the premise outlined by \cite{carr2020option}, this study aims to deepen our understanding of the impact of volatility fluctuations on option portfolios. We have employed the localized volatility model framework to achieve this objective, utilizing the Implied Volatility Surface (IVS). This concept could prove highly advantageous for regulators due to its straightforward implementation while maintaining sufficient accuracy to prevent substantial financial losses.

\begin{figure}
    \centerline{\includegraphics[width=0.9\linewidth]{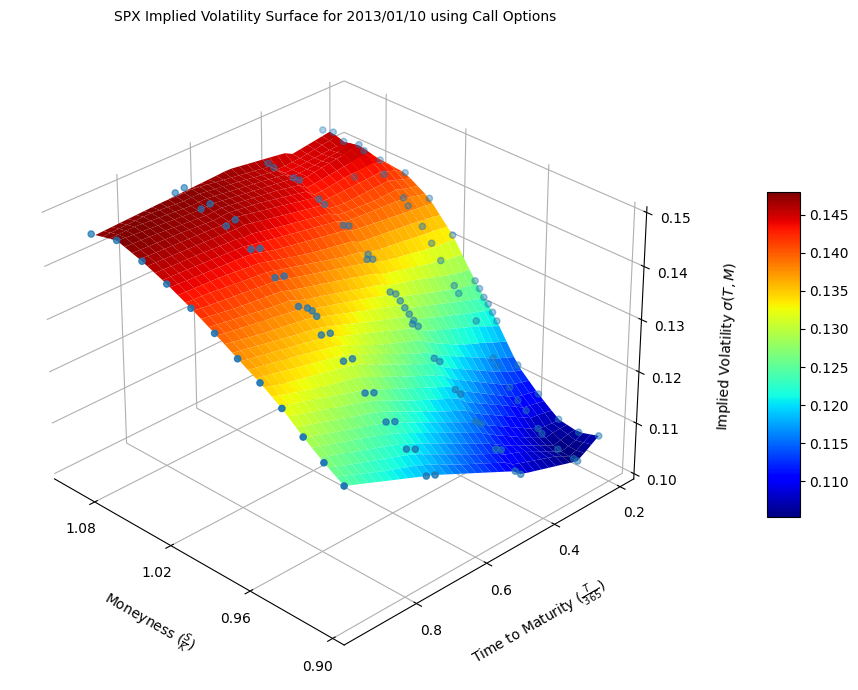}}
    \caption{On January 10, 2013, an examination of implied volatility was conducted across various call options, considering different expiration dates and strike prices. This analysis effectively rejected the hypothesis of constant volatility. The graphical representation of this analysis featured two axes: one is moneyness denoted as $\frac{S}{K}$, where $K$ represents the strike price, and $S$ signifies the underlying asset price. The other is the time-to-maturity of the options, measured in years. Furthermore, the data points available on the volatility surface correspond to the observed implied volatility for the various options under examination.}
    \label{fig:IVS(2013_01_10)}
\end{figure}

\begin{figure}
    \centerline{\includegraphics[width=\linewidth]{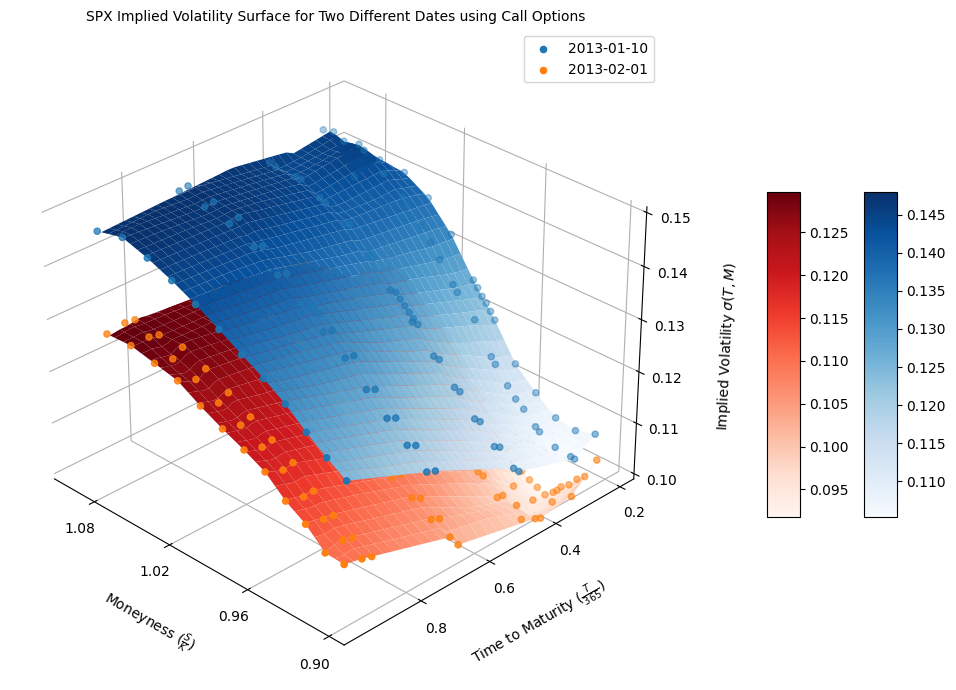}}
    \caption{This Figure presents the Implied Volatility Surface (IVS) of the S\&P 500 index for two distinct dates. The Implied Volatility Surface (IVS) exhibits dynamic characteristics, shifting and altering its shape as the time parameter changes. This dynamism results in a lack of a fixed form for the Implied Volatility Surface (IVS) at any specific moment. Instead, it highlights the inherently uncertain nature of the Implied Volatility Surface (IVS). Again, the data points available on the volatility surface correspond to the observed implied volatility for the various options under examination.}
    \label{fig:IVStwo_dates}
\end{figure}

Implied volatility, a cornerstone of the Black-Scholes pricing model, demonstrates variations across diverse levels of moneyness\footnote{The concept of moneyness in the context of options refers to the current state of an option relative to the prevailing market conditions, with a specific focus on the option's strike price (K) and the present market price of the underlying asset (S). In this study, moneyness for call options is calculated as the ratio $\frac{S}{K}$.} and time-to-maturity, culminating in forming a three-dimensional construct termed the implied volatility surface (IVS), as shown in Figure \ref{fig:IVS(2013_01_10)}. This dynamic surface, analogous to the yield curve (\cite{zamani2022pathwise}), encounters temporal fluctuations (Figure \ref{fig:IVStwo_dates}). Similarities to the yield curve are discernible in the IVS variations, albeit accompanied by the additional complexities introduced by an extra dimension and the variable nature of sample points within the IVS (\cite{skiadopoulos2000dynamics}). Functional representation has emerged as an alternative to point-based methodologies to circumvent challenges tied to sample point adjustments on the IVS. Further intricacies inherent to this phenomenon will be expounded upon in subsequent sections.

We introduce an innovative parametric surface projection methodology, harnessing historical simulation and Monte Carlo techniques. We construct a Profit and Loss (P\&L) distribution by scrutinizing historical observations, enabling the VaR computation. The effectiveness of the proposed approach is subsequently evaluated utilizing data extracted from the U.S. option database\footnote{The data used in this study was sourced from the U.S. option database, accessible at \url{https://optiondata.org/\#sampleId}. It is important to note that only the free data from the website was utilized in the analysis.}. To fulfill this objective, we draw upon established models from the contemporary academic corpus, which have been developed to appraise the methodologies employed for VaR calculations (\cite{roccioletti2015backtesting}).

The dynamic changes observed within the IVS are pivotal factors that significantly impact the pricing dynamics of options contracts. These alterations in IVS are consequential in risk assessment and management within financial markets. These deviations are not unique to a single market but are also evident in the currency market. 

For instance, we can observe similar phenomena in the realm of foreign exchange (FX). The FX implied forward curve is a valuable indicator that reflects market expectations concerning future borrowing rates. Employing parametric projection techniques based on historical scenarios becomes particularly relevant to understand better and leverage these expectations for risk assessment and decision-making.

In summary, the suggested method is not limited to a specific financial domain but extends its influence across various markets, including the currency market. Utilizing parametric projections based on historical scenarios can be a valuable approach to harnessing these dynamics for risk assessment and pricing in these markets.

The subsequent sections of this work are outlined below. Section \ref{sec:Methodology} presents our methodology for estimating the Value at Risk (VaR) arising from IVS movements of option portfolios. In Section \ref{results}, an evaluation of our strategy is conducted utilizing a range of tests and models. The section \ref{conclusion} of the study is brought to a close.
\section{Methodology}\label{sec:Methodology}

This section begins by drawing a parallel between changes in volatility as a risk factor for options portfolios and changes in the interest rate curve for portfolios encompassing bonds. To facilitate this comparison, we will briefly explore the similarities and distinctions between these two phenomena.
\begin{figure*}
    \centerline{\includegraphics[width=\linewidth]{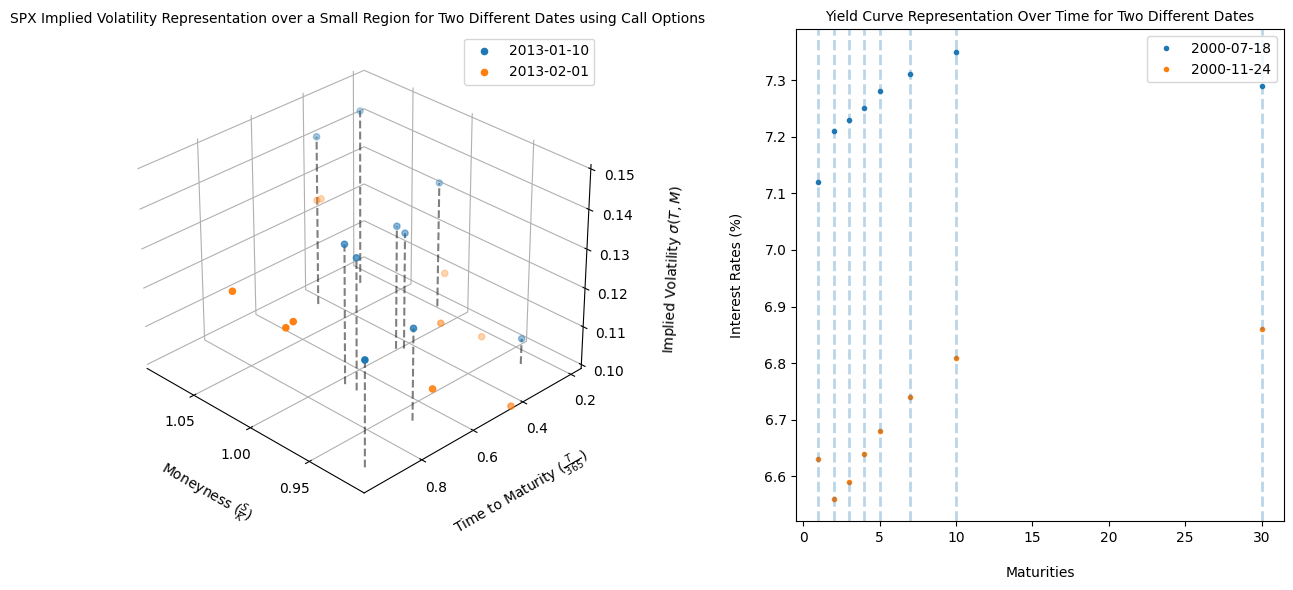}}
    \caption{Sample points of the state variables for the implied volatility surface and the interest rate curve. The right-hand graph illustrates the dynamic fluctuations of the sample points on the interest rate curve, while the left-hand side displays the sample points of the implied volatility surface on two distinct dates.}
    \label{fig:YieldCurve_IVS}
\end{figure*}
\begin{figure}
    \centerline{\includegraphics[width=\linewidth]{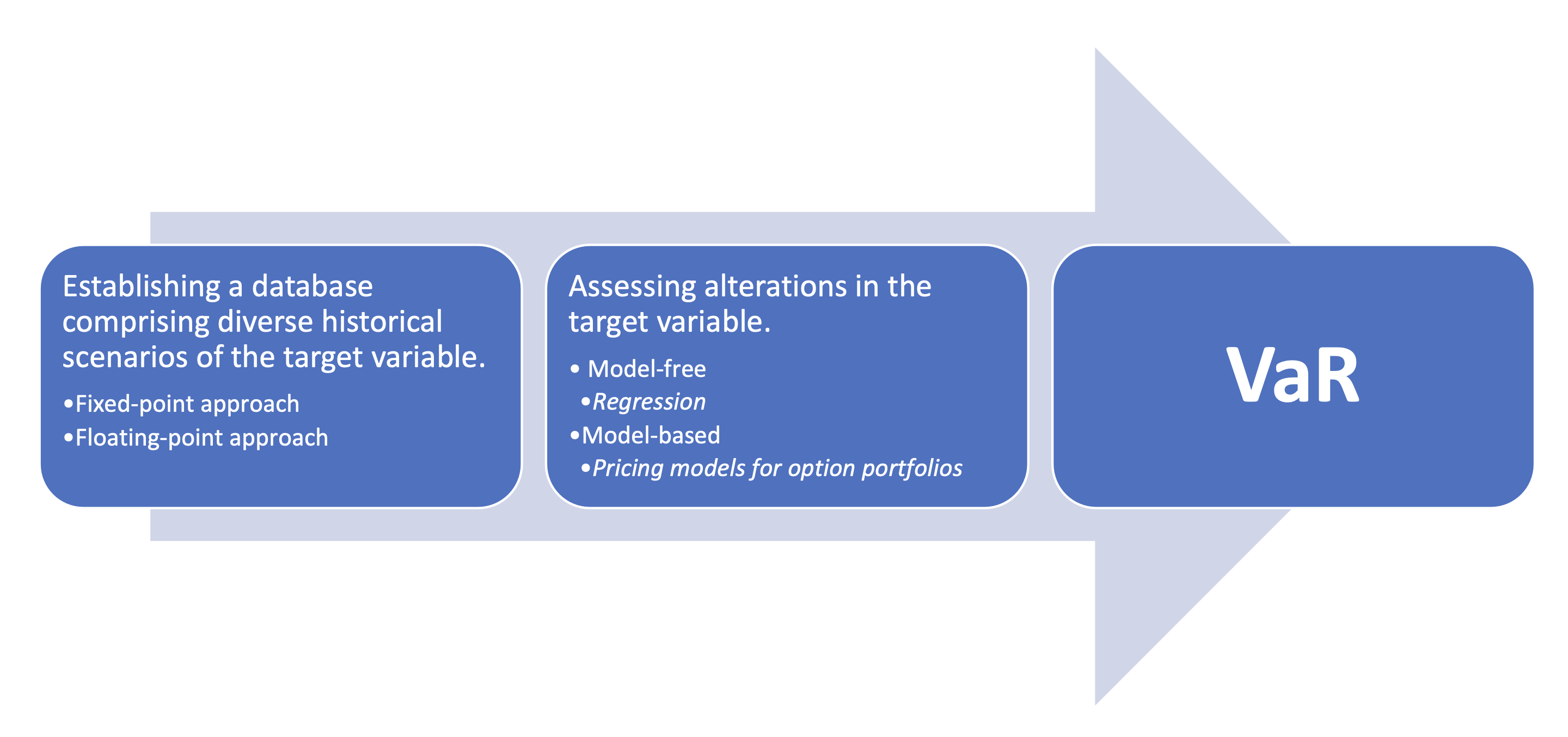}}
    \caption{This diagram illustrates the methodology used to evaluate risk within option and bond portfolios.}
    \label{fig:InterestRate_IVS_models}
\end{figure}
\begin{figure}
    \centerline{\includegraphics[width=\linewidth]{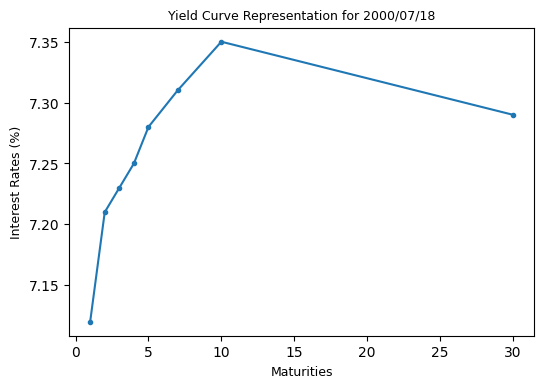}}
    \caption{Swap rate curve of the United States on 07/18/2000}
    \label{fig:YieldCurve(2000_07_18)}
\end{figure}

The process commences with creating a database utilizing historical data, which serves as the foundation for subsequent risk assessment techniques (Figure \ref{fig:InterestRate_IVS_models}). Two distinct approaches are employed to construct this database:
\begin{enumerate}
    \item \textbf{Fixed Point Approach}: This approach is tailored for bond portfolios characterized by fixed-position sample interest rates plotted along the yield curve. In these scenarios, interest rates at predefined points on the yield curve remain constant throughout the assessment. The interest rate curve\footnote{The swap rate data is obtained from the Federal Reserve Economic Data (FRED) online service.}, as depicted in Figure \ref{fig:YieldCurve(2000_07_18)}, is a two-dimensional representation that displays interest rate values corresponding to specific sample points. These sample points are anchored at fixed time intervals, covering maturities ranging from 1 to 30 years. It is worth noting that while this curve evolves over time, the sample points maintain a constant position, as illustrated in the right diagram of Figure \ref{fig:YieldCurve_IVS}.
    \item \textbf{Floating Point Approach}: In contrast, the floating-point approach is applied when dealing with option portfolios. In this context, options are positioned on the implied volatility surface. Unlike bond portfolios with fixed positions, the positions of options on the implied volatility surface are not static but rather dynamic, varying over time (Figure \ref{fig:YieldCurve_IVS}). This characteristic poses a challenge when evaluating the risk associated with fluctuations in the volatility surface. The path of sample points traverses through three dimensions, unlike the interest rate curve, which transitions through a single dimension. Furthermore, apart from predictable changes in the time-to-maturity dimension, alterations in moneyness and implied volatility are inherently stochastic. 
\end{enumerate}

A unique technique is employed to capture historical data for option portfolios, primarily those reliant on implied volatility surfaces. This technique leverages functional representations to transform potential movements' intricate and multidimensional realms into a more manageable set of coefficients. The selection of the degree of freedom determines the precise number of coefficients necessary to represent these historical movements accurately.

This methodology allows for a structured and systematic evaluation of risk within both option and bond portfolios, considering the dynamic nature of implied volatility surfaces in the former and the fixed-position samples along the yield curve in the latter.

Historical scenarios from the first block are transmitted to the second block to assess the impact of changes in target variables on portfolio risk. In this process, two distinct approaches are employed to capture these effects.
\begin{enumerate}
    \item \textbf{Model-free}: In the context of interest-rate models, which aligns with the insights provided by \cite{zamani2022pathwise}, a straightforward technique involves using simple linear regression models. These models are instrumental in quantifying the influence of variable changes on portfolio risk within this domain.
    \item \textbf{Model-based (as applied in this study)}: However, a model-based approach is preferred in more intricate scenarios like option portfolios. Within this study, we focus on option portfolios. Here, model-based techniques are utilized, leveraging option pricing models. These models enable a comprehensive analysis of how variable changes affect the portfolio's risk profile. This approach is precious when dealing with complex financial instruments like options.
\end{enumerate}

Finally, we calculate VaR using the recovered profit and loss distribution from the previous block. As further elucidated in Section \ref{sec:PSP}, to simplify the intricacies of monitoring the dynamics of individual options, we rely on the price changes of the underlying stock as the sole variable necessary for assessing alterations in moneyness and the volatility surface to model changes in volatility for all options. Consequently, the presentation of the volatility surface poses a unique challenge that we address in this section.

\subsection{Volatility surface based on (non)parametric representations}

Implied volatility is typically represented using two distinct methodologies: parametric and non-parametric models. The parametric approach uses specified variables to parameterize implied volatility within a defined domain. A prominent example of this representation can be traced back to \cite{dumas1998implied} study, where a polynomial representation, as depicted in Equation \eqref{regression_model}, is applied to variables encompassing time-to-maturity ($T-t$) and forward moneyness ($M_t$).
\begin{align}
    \sigma^t(M, T) &= \alpha^t_0 + \alpha^t_1M_t + \alpha^t_2M^2_t \nonumber\\
    &+ \alpha^t_3(T-t) + \alpha^t_4M_t(T-t)\label{regression_model}\\
    M_t &= \frac{\ln{F_t} / K}{\sqrt{T-t}}, \; \; \nonumber 
    F_t = S_te^{r(T-t)} \nonumber 
\end{align}
However, a primary limitation of this representation is its failure to guarantee the absence of arbitrage opportunities within the implied volatility surface (\cite{homescu2011implied}). 

Generally, no-arbitrage conditions are studied in two distinct spaces: Static arbitrage and Dynamic arbitrage. Static arbitrage, which centers on immediate opportunities based on current market conditions and does not involve ongoing management, is the most widely recognized. It has been the subject of extensive research, including studies by \cite{roper2010arbitrage} and \cite{niu2015no}. These studies have derived sufficient and nearly necessary conditions for an implied volatility surface to be free from static arbitrage and have identified the conditions under which models fail to guarantee this freedom. On the other hand, dynamic arbitrage entails continuous monitoring and adjustments of options positions to adapt to changing implied volatility levels, thereby maintaining arbitrage opportunities. This form of arbitrage is more intricate and has received less attention in research, with \cite{niu2015no} delving into the conditions under which no-arbitrage is guaranteed. Consequently, further research has been dedicated to developing more refined models.

The SVI (Stochastic Volatility Inspired) parametric model (Equation \eqref{SVI}) distinguishes itself from the polynomial model by its capacity to ensure the absence of arbitrage opportunities across broader ranges (\cite{homescu2011implied}).
\begin{align}
    \sigma^2(x_t) &= a_t + b_t\Big(\rho_t(x_t - m_t) \nonumber \\
    &+ \sqrt{(k_tx_t-m_t)^2+s_t^2}\Big) \label{SVI}
\end{align}
The variable $x_t$ in Equation \eqref{SVI} represents the forward moneyness, while other parameters are determined through optimization over various time periods. As explained in the study by \cite{aurell2014svi}, the model's ability to condition its parameters plays a crucial role in mitigating arbitrage possibilities. \cite{homescu2011implied} evaluations underscore the potential application of this approach within the energy market, demonstrating its versatility across diverse research investigations. Nevertheless, it is essential to note that this approach has inherent complexities in its optimization and calibration processes. Given the daily need to perform this optimization, we employ more simplified approaches.

Most non-parametric approaches rely on the principle of interpolation to estimate the implied volatility surface. Several interpolation methods have been proposed for this purpose; however, it is essential to note that specific methodologies may fail to satisfy the no-arbitrage condition under certain circumstances.

For instance, \cite{clark2011foreign} highlights that using linear interpolation concerning time can lead to impractical predictions of future volatility, potentially giving rise to calendar arbitrage. Therefore, the selection of an appropriate interpolation methodology is of paramount importance.

\cite{homescu2011implied} comprehensive analysis categorizes the various approaches employed in multiple publications into three overarching categories:
\begin{enumerate}
    \item Regular cubic splines
    \item Cubic B-splines
    \item Thin splines
\end{enumerate}
Cubic B-splines are often favored over regular cubic splines due to their enhanced resilience to erroneous data and ability to maintain crucial properties such as monotonicity and convexity (\cite{wolberg2002energy}). Additionally, they offer a higher degree of flexibility, facilitating the creation of diverse shapes. The ability to construct multiple curves, as further elaborated in section \ref{sec:PSP}, allows for generating a broader range of scenarios concerning the dynamics of the volatility surface. 

This study primarily focuses on identifying the optimal B-spline representation for a volatility surface based on observed implied volatility data points. The cubic B-spline interpolation involves breaking down the IVS into a set of cubic polynomial segments defined by a set of control points. These control points act as anchor values for implied volatilities at specific maturities and strikes. The formula for a cubic B-spline segment can be expressed as:
\begin{align}
    S_i(x) = \sum_{j=i-2}^{i+1}\beta_jB_j(x) \label{B-spline}
\end{align}
Where:
\begin{itemize}
    \item $S_i(x)$ represents the cubic B-spline segment between control points $i$ and $i+1$. 
    \item $x$ denotes the strike or moneyness at which implied volatility is estimated.
    \item $B_j(x)$ represents the B-spline basis functions.
    \item $\beta_j$ are the coefficients that determine the shape of the cubic B-spline segment.
\end{itemize}
Constructing an implied volatility surface (IVS) with cubic B-splines involves selecting control points based on market option prices. These control points represent the implied volatilities at specific maturities and moneynesses. The B-spline basis functions, often computed using the Cox-de-Boor recursion formula, are then employed to create cubic B-spline segments between these control points.

These segments smoothly interpolate implied volatilities, ensuring a continuous and differentiable IVS. The entire IVS is constructed by calculating cubic B-spline segments for various moneyness values within the specified maturities. This approach allows for estimating implied volatilities for any moneyness and maturity within the range of control points.

The benefits of this construction process include flexibility in modeling IVS, the creation of smooth and differentiable surfaces, and high precision in estimating implied volatilities, contributing to accurate options pricing and risk management.

In addition, we have applied specific criteria, as outlined in \cite{Verma2008ImpliedVM}, for the interpolation of the volatility surface.

Options with a maturity of fewer than 15 days were excluded from consideration. Ignoring very short-maturity options when constructing an implied volatility surface is typically done for several reasons:
\begin{enumerate}
    \item Very short-maturity options tend to have limited trading activity and liquidity, making them less reliable for estimating implied volatilities.
    \item These options are susceptible to changes in interest rates. They can exhibit erratic price movements due to factors unrelated to the underlying asset's volatility, such as changes in the risk-free rate.
    \item Including very short-maturity options can introduce noise and instability into the implied volatility surface, making it less useful for risk management and pricing purposes.
\end{enumerate}
Therefore, focusing on options with slightly longer maturities, which are more liquid and less affected by interest rate fluctuations, is a common practice when constructing a reliable implied volatility surface.

Options with call prices less than one dollar were likewise omitted from the analysis. Excluding very low-priced options when constructing an implied volatility surface is essential because these options often lack liquidity, have wide bid-ask spreads, and are far out of the money. Their limited sensitivity to market changes and increased model sensitivity can lead to unreliable implied volatility estimates. Additionally, these options may be prone to mispricing and introduce inaccuracies and distortions into the implied volatility surface. Therefore, focusing on liquid and actively traded options enhances the accuracy and reliability of the surface, making it more useful for risk management, pricing, and trading decisions.

The average of the ask and bid price values was utilized to represent the call price. Averaging the bid and ask prices of an option helps determine its market price. This practice is common in financial markets for several reasons. It accounts for buying and selling interest, mitigating the impact of the bid-ask spread. It aligns with market efficiency principles, as efficient markets should reflect actual option values. Averaging also reduces the influence of extreme prices and is considered a fair way to assess market value. Traders often use the mid-price as a reference for executing trades. In options pricing models, the mid-price is a standard input representing a balanced view of supply and demand. However, traders should remain aware of spread variations based on liquidity and market conditions when making trading decisions.

Call prices that fell below the theoretical value, calculated using the BSM model, were disregarded. Excluding options with prices lower than their theoretical BSM prices when constructing an IVS is crucial for several reasons. 

Firstly, low-priced options can introduce noise and inaccuracies into the surface. Such options often have wide bid-ask spreads and limited trading activity, making their prices less reliable market sentiment indicators. By excluding them, the IVS focuses on more liquid and actively traded options, leading to a smoother and more dependable representation of market expectations.

Secondly, very low-priced options may be prone to mispricing or erroneous trades, especially in fast-moving markets. These anomalies can distort the IVS, leading to misleading conclusions about market sentiment. Removing such options helps mitigate these distortions and enhances the surface's accuracy.

Thirdly, including extremely low-priced options on the surface could violate no-arbitrage conditions. The BSM model assumes that options with different strike prices but the same maturity should have consistent implied volatilities. If low-priced options are included without proper screening, this assumption may be violated, opening the door to arbitrage opportunities.

In summary, excluding options with prices lower than their theoretical BSM values when constructing an implied volatility surface ensures the surface's reliability, accuracy, and consistency. It also helps maintain essential no-arbitrage conditions, providing a more robust tool for pricing, risk management, and market analysis.

\subsection{S\&P 500 volatility surface vs. the VIX\protect\footnote{This research is fundamentally grounded in the utilization of the S\&P 500 index. Consequently, every analysis undertaken in this subsection and throughout other sections of this study is contingent upon using this index. The selection of the S\&P 500 index as the core foundation of this research is underpinned by its comprehensive market coverage and well-established status as a benchmark within the market.}}\label{sec:VIX}

Consider a call option with a designated price $C_t$, associated with an underlying security. By the principles of the BSM pricing model, the valuation of this particular option pivots upon the intricate interplay of four pivotal parameters:
\begin{itemize}
    \item $S_t$: The price of the underlying asset.
    \item $T-t$: The temporal gap until maturity.
    \item $\sigma_t$: The anticipated level of volatility.
    \item $r_t$: The prevailing risk-free interest rate.
\end{itemize}
The significance of these parameters, thoughtfully expounded upon within the bounds of Equation \eqref{eq:EC_price}, cannot be overstated. They wield substantial influence over the dynamics governing the pricing of this option. Consequently, forecasting the future prices of $C_{t+1}$ necessitates our capacity to anticipate alterations in each of these fundamental parameters. Given this study's primary thematic thrust on volatility dynamics, our subsequent discourse will delve deeper into this particular parameter.

As illustrated in Figure \ref{chap:intro}, within the hierarchical framework of VaR calculation methods for option portfolios, a distinct bifurcation is observed under the category of \textit{Volatility Exposure}. Herein, two distinct branches emerge, offering divergent methodologies for comprehending the impact of volatility on option portfolio valuations. These methodologies encompass the \textit{Reference Volatility Approach (RVA)} and the \textit{Localized Volatility Approach (LVA)}, each wielding unique insights into the dynamic interplay between volatility and option portfolio pricing dynamics.

In the RVA, all option contracts $i$ are associated with a single reference dynamics specification, where $\sigma_{t}^{(i)}=\sigma_t$ for all $t$. Conversely, the LVA connects the current fair value of an individual option contract's price to variations in the option's implied volatility. Thus, it is not always the case that $\sigma_{t}^{(i)} = \sigma_{t}^{(j)}$ holds for all $i$ and $j$. In this context, both the VIX index and the IVS of the S\&P 500 take center stage as the primary instruments for each respective approach.

The VIX index, often referred to as the \textit{fear gauge} or \textit{fear index}, plays a pivotal role in forecasting imminent volatility levels for the S\&P 500 index. Its primary function is to quantify the anticipated degree of volatility in the S\&P 500 over a defined period, typically spanning the next 30 consecutive calendar days. Elevated VIX values are indicative of an expectation of heightened market volatility, whereas lower values suggest a reduced anticipation of volatility (\cite{whaley2009understanding}). This index, in line with its representation, can be employed as a reference volatility measure to capture the impact of volatility on the price variations of a portfolio.

In the LVA context, IVS plays a crucial role. At the heart of volatility surfaces lies implied volatility, a forward-looking metric that encapsulates market expectations concerning future price volatility (\cite{chan2017machine}). When market participants anticipate increased uncertainty or exhibit apprehension regarding the future performance of an asset, they tend to seek higher premiums for options associated with that asset. This heightened demand results in an increase in the implied volatility of these options. Conversely, during market stability and confidence periods, option prices and implied volatilities tend to decrease.

LVA and RVA differ in various ways. LVA reflects real-time market expectations and is highly dynamic, whereas RVA relies on predefined model assumptions. Option portfolios are typically priced using implied volatility, making LVA more relevant for P\&L attribution. Model discrepancies between LVA and RVA can lead to pricing mismatches, impacting expected P\&L. LVA provides insights into changing market risk perceptions, while RVA may not accurately reflect actual market dynamics. Portfolio managers often base trading strategies on LVA due to its reflection of actual market conditions, whereas RVA is model-dependent and may not capture market nuances effectively (\cite{daglish2007volatility}). In this study, we have employed the LVA with the IVS as our primary tool for capturing local volatility movements, represented as $\Delta\sigma_t^{(i)}=\sigma_{i,t+1}-\sigma_{i,t}$.

\subsection{Risk measurement}\label{sec:VaR}

Value at Risk (VaR) is an extensively utilized risk management tool across various domains. It offers a quantitative metric for assessing potential financial losses incurred by individuals or organizations concerning their investments or financial positions during a specified timeframe, all while upholding a predetermined confidence level. The significance of Value at Risk (VaR) lies in its pivotal role in the evaluation and mitigation of market risk exposure for entities such as financial institutions, investment managers, and traders (\cite{hull2012risk}).

At the heart of the Value at Risk (VaR) concept lies the central inquiry of ascertaining the highest conceivable loss that a portfolio or position might incur over a specified time frame and with a predetermined level of likelihood. Value at Risk (VaR) seeks to approximate the most adverse financial outcome that could transpire under typical market conditions. Mathematically, this concept pertains to the magnitude of the $(1-\alpha)$ percentile within the projected distribution of alterations in baskets, with $\alpha$ representing the confidence interval of VaR. This magnitude is calculated using Equation \eqref{VaR_def}, where the symbol $t$ denotes the current time under consideration, $X_t$ represents the stochastic variable indicating changes in baskets over this period, and $F_{X_t}$ is the cumulative distribution function of $X_t$.
\begin{align}
    \text{VaR}(\alpha, t) = -\text{max}\Big\{x\Big|F_{X_t}(x) \geq 1-\alpha\Big\} \label{VaR_def}
\end{align}
\begin{figure}
    \centerline{\includegraphics[width=\linewidth]{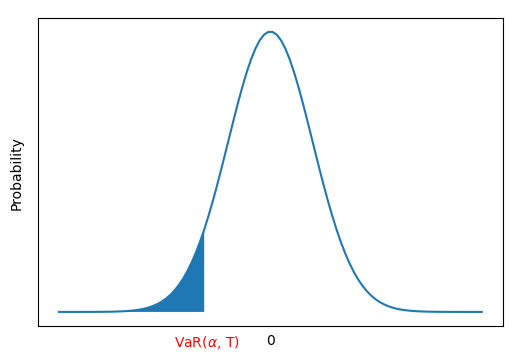}}
    \caption{A representation of the Value-at-Risk (VaR) of a hypothetical distribution}
    \label{fig:VaRdepiction}
\end{figure}
Figure \ref{fig:VaRdepiction} visually represents the Value-at-Risk (VaR) linked with the Profit and Loss (P\&L) distribution. Upon reducing the $\alpha$ value, the blue region within the graph experiences a leftward movement, subsequently leading to an augmentation in the Value-at-Risk (VaR).

VaR is a widely used risk measure but has limitations, including subadditivity and a lack of coherence. These limitations were evident during the 2007-2009 financial crisis when VaR-based regulatory capital calculations underestimated risk. The Basel Committee introduced the Fundamental Review of the Trading Book (FRTB) to address these issues, proposing a shift from VaR to Expected Shortfall (ES) as the primary regulatory capital measure (\cite{hull2012risk}).

ES offers a more comprehensive perspective on risk. Unlike VaR, which only considers the threshold for severe losses, ES also accounts for the number of losses exceeding this threshold (Equation \eqref{ES}). It calculates the average of these severe losses, providing a more nuanced risk assessment.
\begin{align}
    \text{ES}(\alpha, t) = -\mathbb{E}[X_t|X_t \leq q_{1-\alpha}(X_t)] \label{ES}
\end{align}
To reiterate, in this context, $\alpha$ represents the confidence level, $t$ signifies the time horizon, and $X_t$ is the stochastic variable signifying changes in baskets during this timeframe. Furthermore, $q_{1-\alpha}(X_t)$ denotes the ($1-\alpha$)-quantile of the distribution of $X_t$.

However, it is essential to note that VaR remains a standard tool for measuring financial risk in the industry. Moreover, the methods presented in this study can estimate the empirical distribution of portfolio value changes, allowing for the calculation of both VaR and ES. This approach ensures flexibility in risk measurement, accommodating the strengths of both measures.

Multiple approaches are utilized in the calculation of VaR. Determining VaR within the parametric framework (\cite{hull2012risk}) entails using a specified probability distribution, such as the normal distribution, to represent profit and loss. The VaR calculation is derived from the underlying distribution, denoted as $F_{X_t}$, as expressed in Equation \eqref{VaR_def}. An inherent limitation linked with calculating VaR through the parametric approach is the reliance on the normality assumption. Financial markets are susceptible to heightened volatility and non-normal patterns, leading to potentially inaccurate estimations, especially during market crises. Hence, prudent consideration of alternative methodologies becomes imperative.

Historical simulation (HS) is a prevalent and straightforward approach frequently employed in academic research (\cite{hendricks1996evaluation}). This methodology involves extracting feasible scenarios for modifying a portfolio's value by leveraging historical data. The subsequent calculation of VaR involves evaluating the utmost potential loss sustained by the portfolio, factoring in all historical losses while adhering to a designated confidence level. However, a notable limitation inherent to this approach stems from its retrospective nature. The dynamism of market conditions does not comprehensively account for future outcomes.

The Monte Carlo (MC) method is a widely recognized approach employed for calculating VaR. Using a predetermined probability distribution, this technique generates potential outcomes by simulating stochastic future returns (\cite{hull2012risk}). Subsequently, calculating VaR entails pinpointing the loss percentile within the simulated distribution. This methodology proves advantageous when grappling with intricate portfolios harboring non-linear relationships. However, akin to preceding methods, the MC approach also grapples with limitations, notably the inability to identify Black Swan events and the subsequent costs incurred during simulation (\cite{hong2014monte}).

As outlined in \ref{sec:PSP}, our proposed methodology integrates the MC and HS techniques for VaR estimation. This approach holds several inherent advantages, which we will elaborate upon in subsequent deliberations.

\subsection{Parametric Surface Projection (PSP) PnL calculation}\label{sec:PSP}

\begin{figure*}
    \centerline{\includegraphics[width=\linewidth]{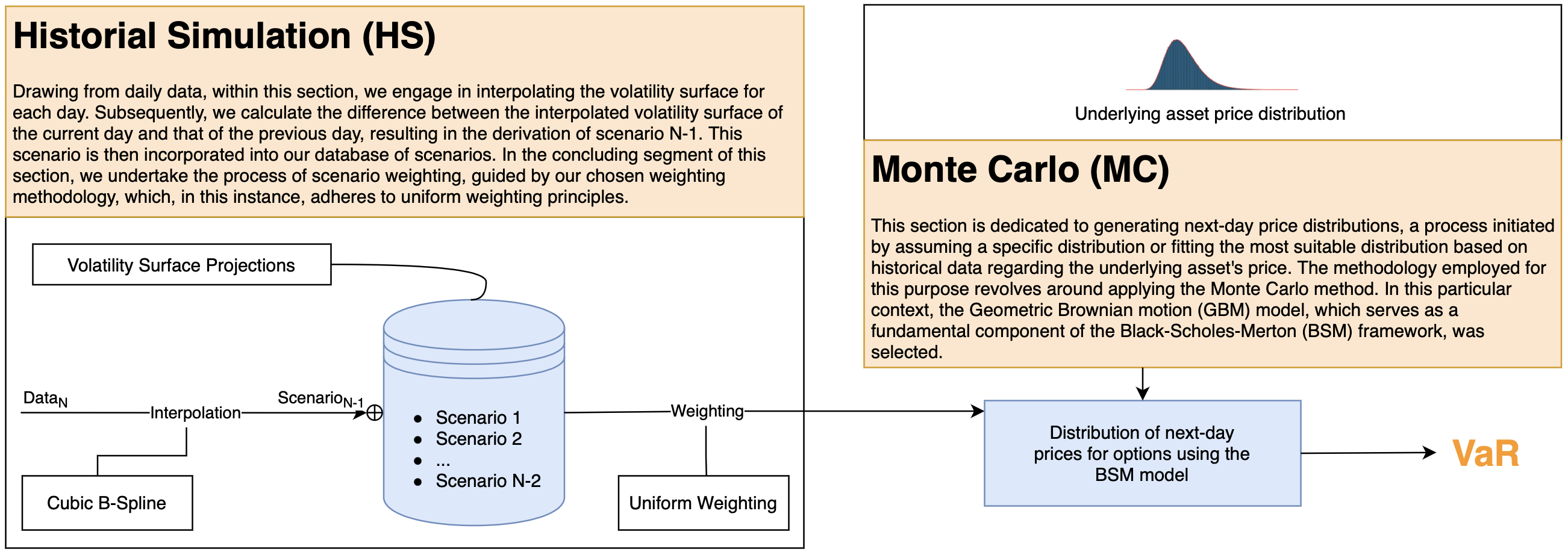}}
    \caption{The flowchart of the Parametric Surface Projection (PSP) method is structured into two principal components. In the first segment, we leverage the Historical Simulation (HS) method to derive an empirical distribution that characterizes the movements within the volatility surface. This distribution enables the projection of potential movements for the forthcoming day. The second segment of the flowchart integrates the Monte Carlo (MC) method, which generates a distribution encompassing the underlying asset's price. We obtain the next-day option price distribution by combining the HS method's empirical and MC-generated distribution. Subsequently, this distribution facilitates the Value-at-Risk (VaR) calculation.}
    \label{fig:Model}
\end{figure*}

This section offers a comprehensive guide to the essential steps in calculating option portfolios' one-day Value-at-Risk (VaR), as illustrated in Figure \ref{fig:Model}. It is vital to revisit Equation \eqref{eq:EC_price}, which derives the call option price through the Black-Scholes-Merton (BSM) model. According to this pricing model, the primary determinants of the option price are time-to-maturity ($T-t$), expected volatility ($\sigma_t$), underlying asset price ($S_t$), and interest rate ($r_t$). Moreover, it is essential to clarify that $K$ represents the strike price associated with an option, and $N()$ corresponds to the cumulative density function of the normal distribution.
\begin{align}
    C_t &= N(d_1)S_t - N(d_2)Ke^{-r_t(T-t)} \label{eq:EC_price}\\
    d_1 &= \frac{1}{\sigma_t \sqrt{T-t}} \Big[ln\Big(\frac{S_t}{K}\Big) + \Big(r_t+\frac{\sigma_t^2}{2}\Big)(T-t)\Big] \nonumber \\
    d_2 &= d_1 - \sigma_t \sqrt{T-t} \nonumber
\end{align}

Time-to-maturity is a deterministic variable, making its next-day value straightforward to determine. Similarly, during the periods leading up to the first half of 2013\footnote{For data related to interest rates and their associated volatility, we accessed the Federal Reserve Economic Data (FRED) platform, available at \url{https://fred.stlouisfed.org/series/FEDFUNDS}.}, interest rates were relatively stable due to their limited, low-frequency fluctuations. Additionally, given our primary focus on short-term effects, it is reasonable to assume that the next day's interest rate remains constant. However, alternative methods, as detailed in \cite{zamani2022pathwise}, can be employed for those who require more precise estimations, particularly in periods of heightened interest rate volatility.

However, the situation becomes notably more complex when dealing with variables like the underlying asset price and expected volatility. These variables are inherently stochastic and subject to frequent fluctuations, demanding specialized techniques for estimation.

In this context, it is imperative to encompass all conceivable trajectories and potential next-day perturbations concerning expected volatility and underlying asset price. When expected volatility exhibits stochastic variations over time, evaluating the extent of this variability and its co-movement with fluctuations in security prices becomes essential.
\begin{figure}
    \centerline{\includegraphics[width=\linewidth]{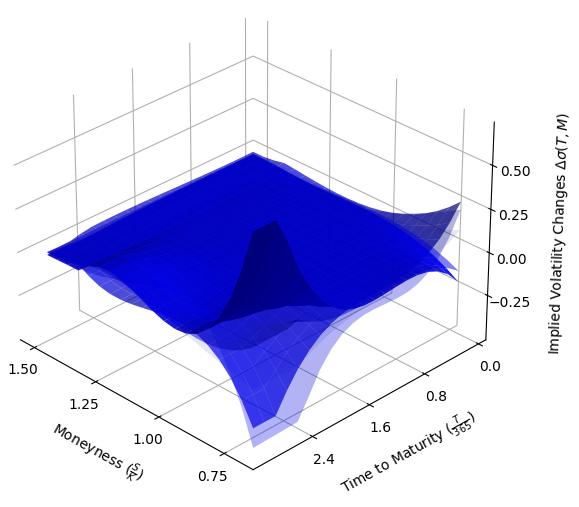}}
    \caption{Five random scenarios of Implied Volatility Surface (IVS) movements are depicted. The surfaces are color-coded, with lighter colors indicating older scenarios and gradually transitioning to darker blue shades as we progress to more recent dates.}
    \label{fig:IVS_movements}
\end{figure}

The MC component embedded within the PSP method simulates next-day security prices. This simulation process facilitates estimating an empirical distribution for expected volatility and the computation of next-day option prices. The selected model for simulating next-day security prices in this study is the Geometric Brownian motion (GBM) model, which is a fundamental component of the BSM framework.
\begin{align}
    dS_t = \mu S_tdt + \sigma S_t dB_t \label{GBM}
\end{align}
Here, $B_t$ signifies a standard Wiener process, with $\mathbb{E}[dB_t]=0$ and $V[dB_t]=dt$. The parameters $\mu$ and $\sigma$ denote the instantaneous drift and volatility of the stock's returns. 

The instantaneous drift term, $\mu$, represents the risk-free interest rate ($r$) minus the underlying asset's continuous dividend yield ($q$). It can be expressed as $r - q$. In addition to the instantaneous drift term, denoted as $\mu$, other parameters, such as $\sigma$, are also assumed to be constant. This assumption is made due to the absence of a closed-form solution for time-variant versions of these parameters. Consequently, for the sake of simplicity, it is assumed that these parameters remain constant over the selected time horizon. Furthermore, their values are set to their daily averages from a year before the starting date of the data, providing a stable foundation for the simulations. It is worth noting that the number of samples can be set judiciously, striking a balance that avoids significant underestimations while refraining from excessive use of far-past samples, which may not accurately represent the underlying distribution, especially when it has undergone substantial changes. The parameters $r$ and $q$, which denote the daily averages mentioned earlier, are approximated as follows: $r = 14.06\%$ and $q = 1.94\%$\footnote{The yearly average of the S\&P 500 dividend yield was sourced from the following link: \url{https://www.multpl.com/s-p-500-dividend-yield/table/by-year}}.
\begin{align}
    S_t = S_0 \text{exp}\Big(\Big(\mu - \frac{\sigma^2}{2}\Big)(t - t_0) + \sigma(B_t - B_{t_0})\Big) \label{stockprice}
\end{align}

Equation \eqref{stockprice} is derived by integrating Equation \eqref{GBM} over the time interval $[t_0, t]$, yielding the stock price at time $t$. Here, $S_0$ denotes the known stock price at the initial time $t_0$. Now, we discretize Equation \eqref{stockprice} to make it suitable for simulations. The expression for the stock price at day $t+1$, given the stock price at day $t$, is as follows:
\begin{align}
    S_{t+1} = S_t \text{exp}\Big(\mu - \frac{\sigma^2}{2} + \sigma\epsilon_t\Big) \label{stockprice_discrete}
\end{align}
Where $\epsilon_t \simiid N(0,1)$ and $\sigma \approx 0.05$ using historical averages mentioned earlier. Ultimately, by generating 1000 pseudo-random numbers at each time step $t$, we simulate the stock price at day $t+1$, denoted as $S_{t+1}$.

The Historical Simulation (HS) component within the PSP method, as depicted in Figure \ref{fig:IVS_movements}, plays a pivotal role in capturing the entirety of plausible movements within the volatility surface, drawing insights from historical data to predict expected volatilities for the forthcoming day. Moreover, it captures the dependency of expected volatilities on security price movements, a crucial factor in determining next-day moneyness, which, in turn, informs the valuation of expected volatilities.

Alternative methods for calculating the dependence between volatility and stock price movements exist. In our approach, we utilize a functional representation of the volatility surface to capture these dependencies. However, model-free techniques, such as the method of moments, can also be applied. For instance, one could focus on first-order shocks, which involve correlated shocks to both the stock price and implied volatility. If we examine the discrete form of GBM using one-day changes, as shown in Equation \eqref{GBM_discrete}, we can also apply a similar assumption to the volatility term, leading to Equation \eqref{GBM_discrete_sigma}. 
\begin{align}
    \Delta S_t &= \mu S_t + \sigma_{t}^{(i)} S_t \epsilon_{t, i} \label{GBM_discrete}\\
    \Delta \sigma_{t}^{(i)} &= \hat{\mu} \sigma_{t}^{(i)} + \widehat{\sigma_i} \sigma_{t}^{(i)} \widehat{\epsilon_{t, i}} \label{GBM_discrete_sigma}
\end{align}
Where $\epsilon_{t, i}, \widehat{\epsilon_{t, i}} \simiid N(0,1)$, but $\epsilon_{t, i} \independent \widehat{\epsilon_{t, i}}$ does not necessarily holds. The parameter $\widehat{\sigma_i}$ represents the volatility of implied volatility, which characterizes the volatility of the volatility itself. The covariance between the returns on volatility ($\widehat{r_{t, i}}$) and the stock price ($r_t$) can be expressed as:
\begin{align}
    \text{cov}(\widehat{r_{t, i}}, r_t) &= \text{cov}\Big(\frac{\Delta S_t}{S_t}, \frac{\Delta \sigma_{t}^{(i)}}{\sigma_{t}^{(i)}}\Big) \nonumber\\
    &= \text{cov}(\sigma_{t}^{(i)}\epsilon_{t, i}, \widehat{\sigma_i}\widehat{\epsilon_{t, i}}) \label{cov_sigma_s(1)}
\end{align}
Since $\sigma_{t}^{(i)} \independent \epsilon_{t, i}, \widehat{\epsilon_{t, i}}$ Equation \eqref{cov_sigma_s(1)} simplifies to:
\begin{align}
    \text{cov}(\widehat{r_{t, i}}, r_t) &= \sigma_{t}^{(i)}\widehat{\sigma_i}\text{cov}(\epsilon_{t, i}, \widehat{\epsilon_{t, i}}) \label{cov_sigma_s(2)}
\end{align}
and consequently,
\begin{align}
    \text{cov}(\epsilon_{t, i}, \widehat{\epsilon_{t, i}}) &= \text{corr}(\widehat{r_{t, i}}, r_t) = \rho_{t,i}
\end{align}
With the obtained correlation between each implied volatility and the underlying price, and given that their distribution of residuals follows a normal distribution, we can proceed to generate random shocks using MC simulation. A similar approach can be employed to model simultaneous shocks to the implied volatility of individual options, capturing their co-movements. However, this process is more intricate, requiring Cholesky decomposition of the correlation matrix of volatility returns to generate pseudo-random numbers. However, employing the HS method through the volatility surface captures dependencies between volatilities and between volatilities and the stock price, primarily due to the inherent structural properties of the surface. Achieving this level of complexity might be more challenging when using the method of moments, as demonstrated earlier. We will now provide a more in-depth explanation of the HS module.

We consider the set $\mathcal{D}^{(n)} = \{\Delta \textit{IVS}_1, \Delta \textit{IVS}_2, \ldots, \Delta \textit{IVS}_{n-1}\}$, encompassing an array of all conceivable scenarios of volatility movements on day $n$. We initiate our scenario generation process on the third day since we require data from the first two days to create at least one scenario for the volatility surface on the subsequent day. From day three onwards, the number of scenarios systematically grows. For instance, on the final day, day 124, we have 122 scenarios. Next, We derive an empirical distribution characterizing expected volatility within each scenario denoted as $\Delta \textit{IVS}_j$. This is achieved by substituting a specific set of variables from Equations \eqref{eq:next-moneyness} and \eqref{eq:next-IVS} into Equation \eqref{eq:next-vol}.
\begin{align}
    m_{t+1} &= \frac{S_{t+1}}{K} \label{eq:next-moneyness}\\
    \textit{IVS}_{t+1} &= \textit{IVS}_t + \Delta \textit{IVS}_j \label{eq:next-IVS}\\
    \sigma_{t+1}^{(i)} &= \textit{IVS}_{t+1}(m_{t+1}, T-t-1) \label{eq:next-vol}
\end{align}
In Equation \eqref{eq:next-moneyness}, we deduce the next-day moneyness by leveraging the distribution of next-day prices of the underlying asset derived from the Monte Carlo (MC) block. Subsequently, for scenario $i$, we construct the next-day volatility surface, as depicted in Equation \eqref{eq:next-IVS}. Within this framework, Equation \eqref{eq:next-vol} plays a pivotal role in computing the distribution that characterizes the next-day expected volatility of scenario $i$.

Furthermore, our analysis delves into the derivation of the next-day volatility surface, a facet achieved by employing historical scenarios as depicted in Equation \eqref{eq:next-IVS}. In this context, we determine the expected volatility distribution affiliated with each scenario. All distributions are amalgamated through a judicious weighting procedure to culminate the process, a step exemplified in Equation \eqref{eq:next-vol-agg}. Uniform weighting has been employed for this specific instance, though it is essential to acknowledge the potential for alternative weighting approaches.
\begin{align}
    f_{\sigma_{t+1}}(\sigma_{t+1} = \sigma) &= \sum_{i}P\Big[\sigma_{t+1} \in \sigma_{t+1}^{(i)}\Big] \nonumber\\ 
    &\times f_{\sigma_{t+1}^{(i)}}\Big(\sigma_{t+1}^{(i)} = \sigma\Big)\label{eq:next-vol-agg}
\end{align}
Each $\tiny P\Big[\sigma_{t+1} \in \sigma_{t+1}^{(i)}\Big]$ represents the probability of an event, where $\tiny\sum_{i}P\Big[\sigma_{t+1} \in \sigma_{t+1}^{(i)}\Big]=1$.

In this specific example, historical data from the more distant past remains relevant and applicable to the current situation. Therefore, uniform weighting may be the more appropriate choice. This relevance is attributed to the specific time horizon, which pertains to a short period in 2013 when market turbulence was at its minimum. However, one may assign greater weight to scenarios representing severe market conditions, a strategy particularly valuable during periods of significant turbulence for stress testing. Additionally, if recent data is a better indicator of current market dynamics, exponential weighting could be a better choice.

In the final phase of our analysis, we compute the distribution for each option's price for the forthcoming day. This calculation hinges on the availability of the next-day distribution encompassing expected volatility ($\sigma_{t+1}$) and the underlying asset price ($S_{t+1}$). Furthermore, we incorporate the assumption of a constant interest rate spanning a single day while simultaneously advancing the time-to-maturity by one step. Calculating the distribution for each option price is executed by substituting pertinent variables into Equation \eqref{eq:EC_price}.

\subsection{Alternative models: Const. Vol \& VIX}\label{sec:alternatives}

Two additional models, Const. Vol and VIX are employed to compare results with the PSP method. In the PSP method, the distribution of next-day prices is derived from Phrase \eqref{PSP-updates}, incorporating updates in local volatility.
\begin{equation}
    C_{t+1}^{(i)}(S_t, \sigma_{t+1}^{(i)}, r_t, T-t, K) \label{PSP-updates}
\end{equation}
For the Const. Vol model, we consider partially constant volatility, where volatility remains unchanged from one day to another, as shown in Phrase \eqref{Const.Vol-updates}.
\begin{equation}
    C_{t+1}^{(i)}(S_t, \sigma_{t}^{(i)}, r_t, T-t, K) \label{Const.Vol-updates}
\end{equation}
Lastly, the reference volatility model is utilized in the VIX model, where volatility updates are consistent across all option prices. This model uses historical simulation from changes in the VIX index instead of volatility surface shocks while keeping all other factors constant. Phrase \eqref{VIX-updates} illustrates the mathematical formula governing the computation of next-day prices in this model.
\begin{equation}
    C_{t+1}^{(i)}(S_t, \sigma_{t}, r_t, T-t, K) \label{VIX-updates}
\end{equation}
\section{Empirical Results}\label{results}

In this section, we proceed to present our findings. We have adopted a data window spanning 122 days to estimate Value-at-Risks (VaRs). Subsequently, we compare these estimations with the realized returns to assess the effectiveness of our methodology. To evaluate our approach, we have formulated a random portfolio comprising 100 call options and employed a buy-and-hold strategy until the options reach maturity. While alternative strategies are available, we have opted for this one due to its simplicity. The data on the options covers the period from January 3rd to June 27th of 2013. However, it is essential to note that other time frames can also be considered. Our portfolio has been meticulously designed to account for various moneyness shifts across different maturity dates. 

\begin{figure*}
    \centerline{\includegraphics[width=\linewidth]{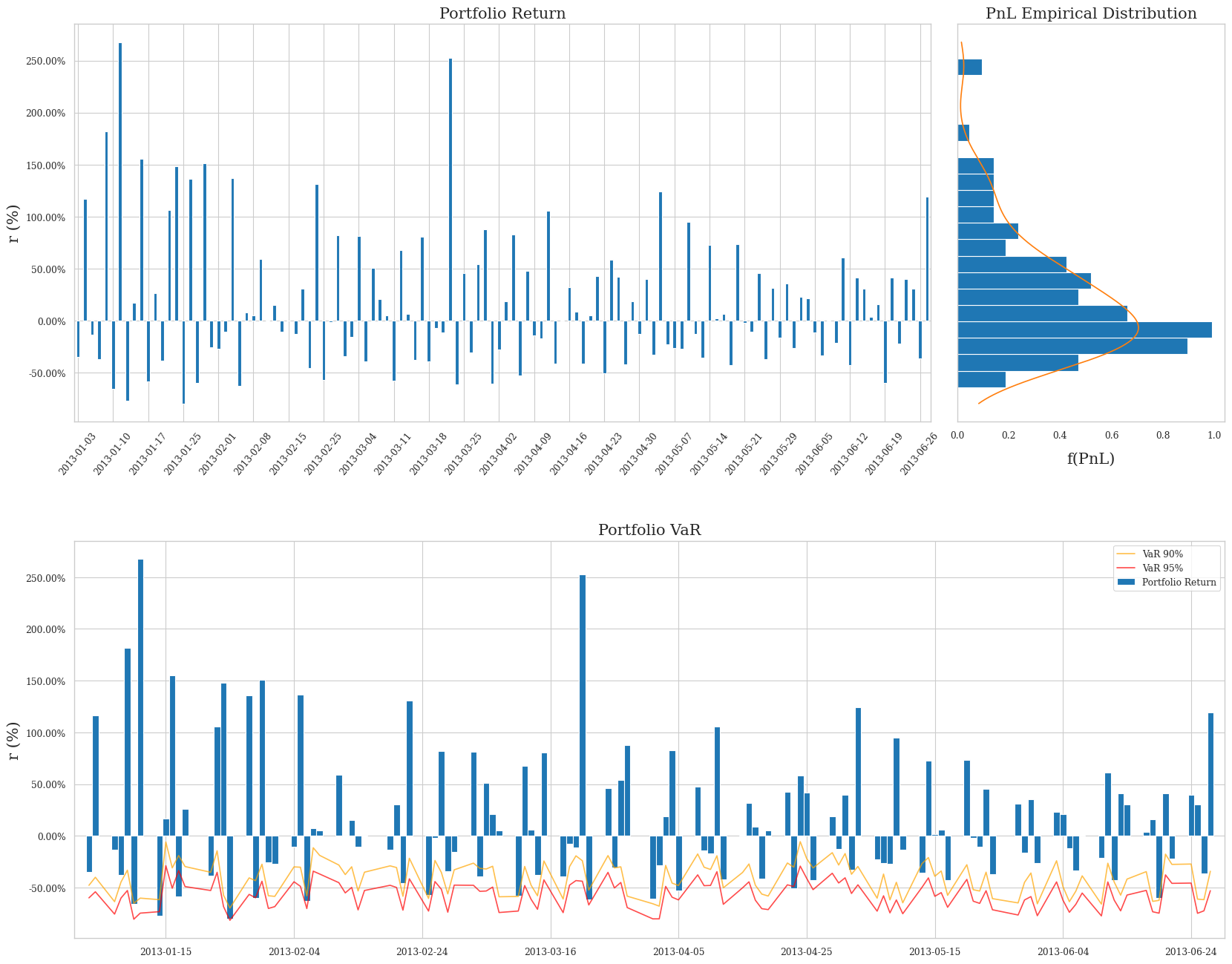}}
    \caption{This figure illustrates the proposed hierarchy of VaR calculation methods within option portfolios, as introduced in the introduction. It is important to note that the hierarchy does not imply any temporal precedence of one method over another. Instead, it visually represents how our suggested model relates to other methods in this context.}
    \label{fig:Portfolio_PnL_Total}
\end{figure*}

\begin{figure}[pos=ht!]
    \centerline{\includegraphics[width=\linewidth]{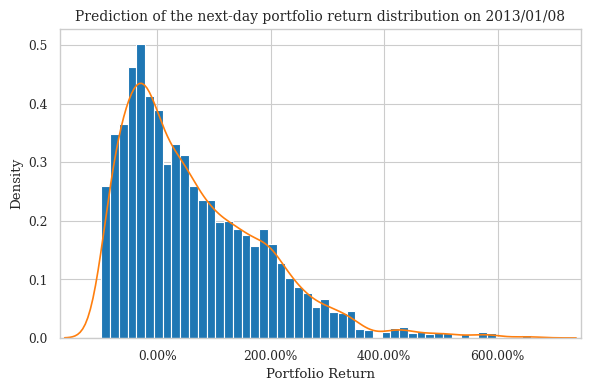}}
    \caption{This figure illustrates a sample of the generated distribution of Profit and Loss (PnL) for a randomly selected date using the PSP method.}
    \label{fig:Portfolio_PnL_day4}
\end{figure}

In Figure \ref{fig:Portfolio_PnL_Total}, the top bar plot illustrates the portfolio's return over the studied period, while its distribution of returns is depicted to the right. It is worth noting that the returns distribution is right-skewed, as the downside is limited to -100\%. At the bottom of the figure, the 90\% and 95\% VaR are outlined using the PSP method. It is important to note that a 99\% confidence level is not included in our analysis due to insufficient data. Figure \ref{fig:Portfolio_PnL_day4} illustrates a sample of the generated distribution of PnL for a randomly selected date employing the PSP method\footnote{The code was executed on an Apple M1 processor, requiring approximately 12,500 seconds for completion.}. 

\subsection{Statistical tests to evaluate Value-at-Risk (VaR) Models}\label{sec:VaR-backtesting}

The first test is the unconditional coverage test, which evaluates the efficacy of a Value-at-Risk (VaR) calculation technique by examining the frequency of its breaches concerning the specified VaR confidence level. The second test, the independence test (\cite{roccioletti2015backtesting}), assesses whether the VaR violations demonstrate independent clustering or dispersion patterns across time. Finally, the conditional coverage test integrates components from the preceding two tests. These tests independently evaluate the dependability of each method and do not entail direct comparisons among them. Acknowledging that rejecting the null hypothesis in these tests indicates the insufficiency of a particular approach is crucial.

\begin{figure*}
    \centerline{\includegraphics[width=\linewidth]{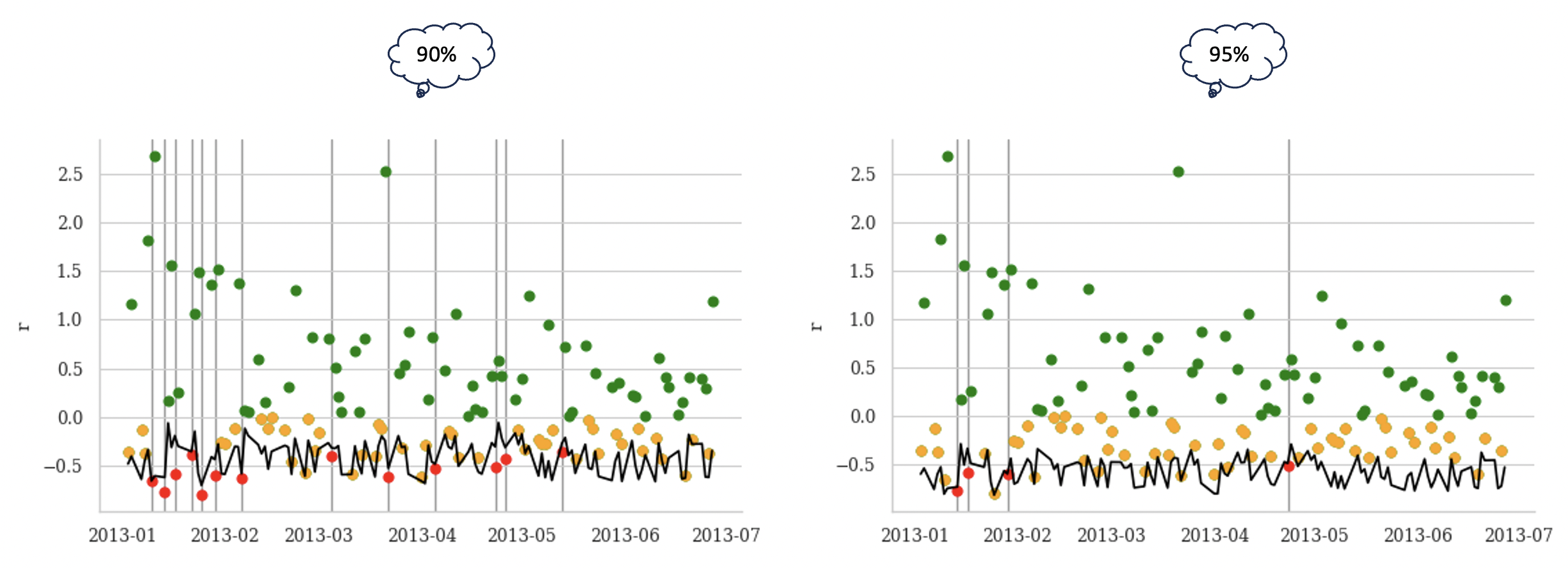}}
    \caption{This figure illustrates various confidence levels associated with VaR calculated using the PSP method. The dots represent portfolio returns for each day, with green indicating positive returns, yellow for negative returns, and red for returns exceeding VaR levels.}
    \label{fig:PSP_R1}
\end{figure*}
\begin{table*}
\begin{center}
\begin{tabular}{*9c}
\hline 
\hline
\toprule
Confidence & Method & \multicolumn{2}{p{2cm}}{\centering Conditional Coverage \\ Test} & \multicolumn{2}{p{2cm}}{\centering Independence \\ Test} & \multicolumn{2}{p{2cm}}{\centering Unconditional Coverage \\ Test} & \multicolumn{1}{p{1.2cm}}{\centering Violation \\ Rate} \\
\cline{3-8}
           &        & Stat.        & P-Value & Stat.& P-Value & Stat. & P-Value &   \\
\hline
90\%    & PSP    & 3.21    & 0.2009 & 3.14  & 0.0765 & 0.07  & 0.7873 & 0.1066    \\
    & Const. Vol & 2.87    & 0.2386 & 1.13  & 0.2870 & 1.73  & 0.1881 & 0.0656    \\
    & VIX        & 3.63    & 0.1626 & 0.86  & 0.3537 & 2.77  & 0.0959 & 0.0574    \\
\hline
95\%    & PSP    & 1.10    & 0.5770 & 0.27  & 0.6010 & 0.83  & 0.3634 & 0.0328    \\
    & Const. Vol & 6.74    & 0.0344 & 0.02  & 0.8973 & 6.72  & 0.0096 & 0.0082    \\
    & VIX        & 6.74    & 0.0344 & 0.02  & 0.8973 & 6.72  & 0.0096 & 0.0082    \\
\bottomrule
\hline
\hline
\end{tabular}
\caption{The performance of the PSP, Const. Vol and VIX methods in estimating VaR with 90\% and 95\% confidence levels are assessed through conditional coverage, unconditional coverage, and independence tests.}
\label{tab:results_tests}
\end{center}
\end{table*}
Additionally, we employed the Diebold and Mariano (DM) test (\cite{epftoolbox}) to conduct pairwise comparisons of the accuracy between different time series models. This test calculates the difference between the forecasted and actual values for each time series, utilizing a loss function. In our case, the Mean Squared Error was employed as the chosen loss function. Subsequently, a statistical test is applied to determine whether the performance of one model significantly outperforms the other.

\begin{figure*}
    \centerline{\includegraphics[width=\linewidth]{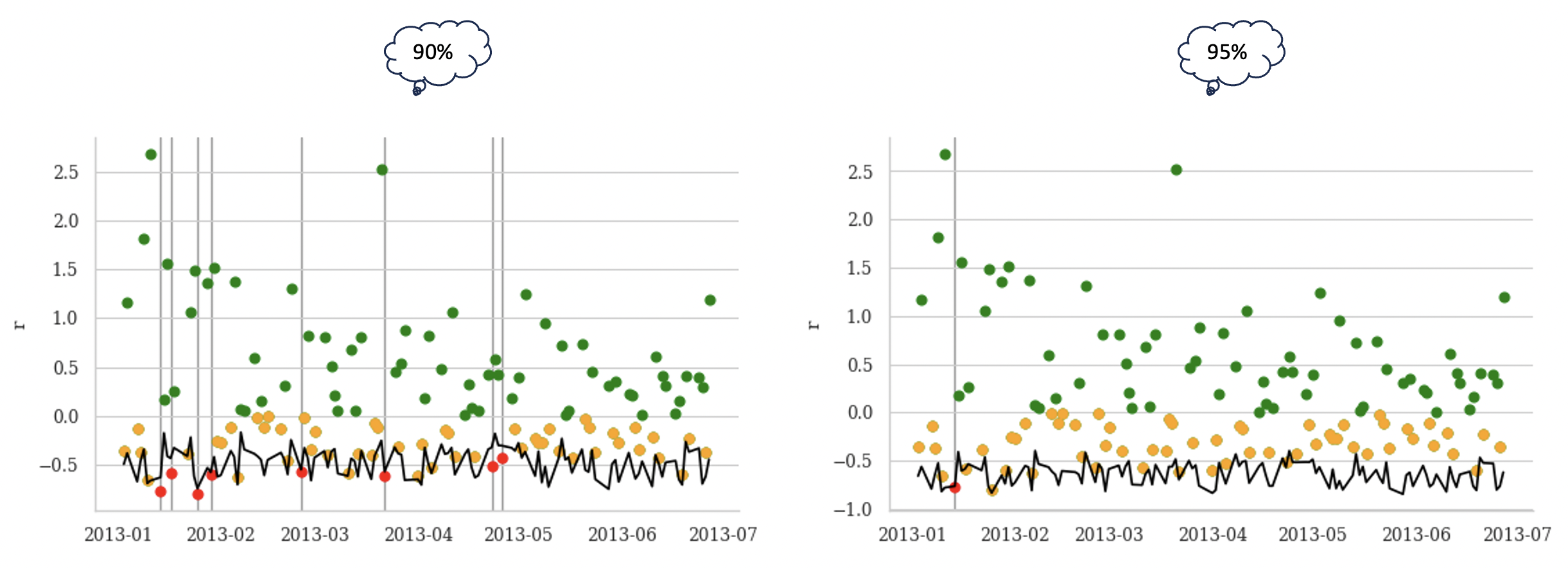}}
    \caption{This figure illustrates various confidence levels associated with VaR calculated using the Const. Vol method. The dots represent portfolio returns for each day, with green indicating positive returns, yellow for negative returns, and red for returns exceeding VaR levels.}
    \label{fig:Const_Vol_R1}
\end{figure*}

\begin{figure*}
    \centerline{\includegraphics[width=\linewidth]{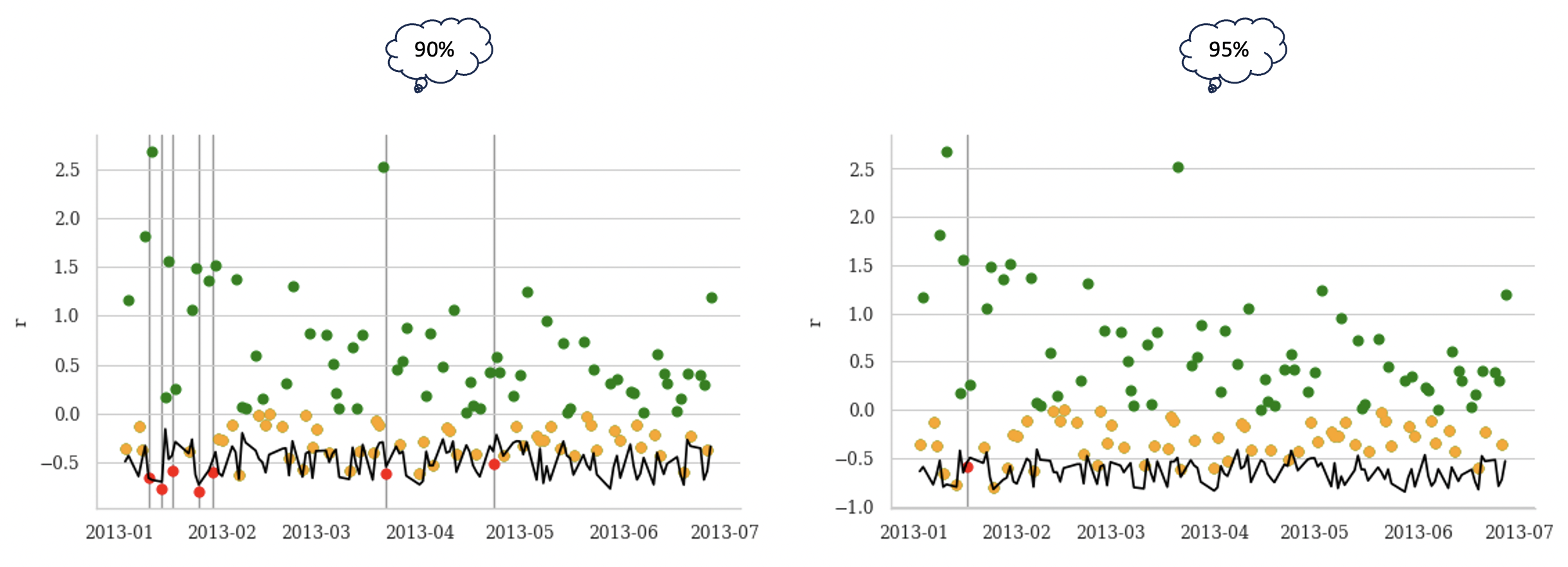}}
    \caption{This figure illustrates various confidence levels associated with VaR calculated using the VIX method. The dots represent portfolio returns for each day, with green indicating positive returns, yellow for negative returns, and red for returns exceeding VaR levels.}
    \label{fig:VIX_R1}
\end{figure*}

Before delving into tables and statistical analysis, we examine the VaR calculations of different methods, illustrated in Figures\footnote{These figures are generated utilizing the GitHub repository available at the following link: \url{https://github.com/BayerSe/VaR-Backtesting}} \ref{fig:PSP_R1}, \ref{fig:Const_Vol_R1}, and \ref{fig:VIX_R1}. Each figure portrays two plots for each confidence interval. Colored dots represent realized portfolio returns: green for positive returns, yellow for negative, and red when the return surpasses the VaR calculated in black. 

These figures reaffirm the superiority of the PSP method. At the same time, violations are infrequently observed in the Const. Vol and VIX models, the PSP method exhibits a notable number of violations, enhancing the reliability of our predictions. Predictions generated by the Const Vol and VIX models tend to be more conservative than those produced by the PSP method. Consequently, these models might suit regulators, emphasizing prudent capital allocation. However, the PSP method stands out for firms seeking optimal capital utilization. The ranking model will enable a comprehensive assessment, considering both regulatory and firm-oriented objectives, for evaluating the accuracies of different models.

Table \ref{tab:results_tests} presents the outcomes of unconditional coverage, independence, and conditional coverage tests for the PSP, Const. Vol, and VIX methods. Analyzing the statistics and violation rates reveals comparable performance among all methods in a 90\% confidence interval, with the PSP method demonstrating a slightly superior violation rate. However, the PSP method outshines the other two at a higher confidence level of 95\%, highlighting its effectiveness. 
\begin{figure*}
    \centerline{\includegraphics[width=\linewidth]{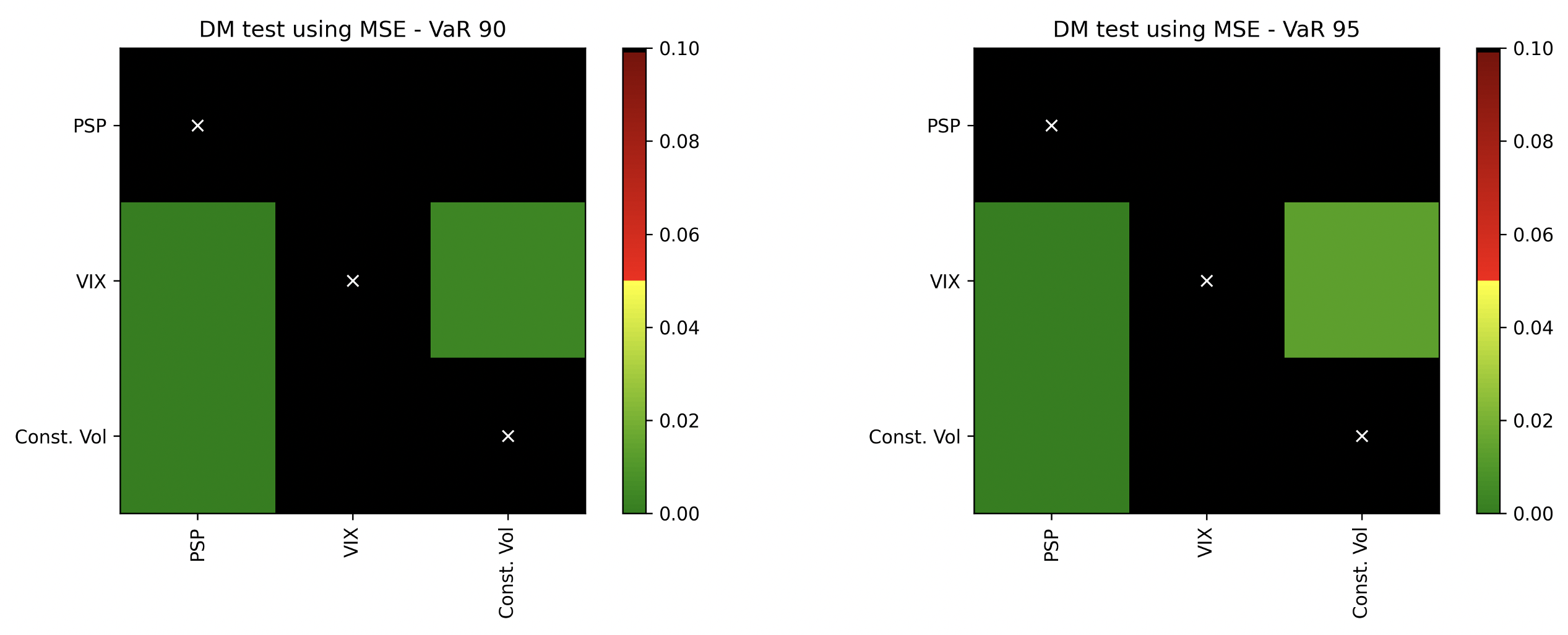}}
    \caption{This figure showcases a chessboard-shaped heat map, presenting the outcomes of the Diebold and Mariano (DM) test for comparing forecasts from various models. The map reveals the p-values associated with the null hypothesis that the forecast on the y-axis is significantly more accurate than the forecast on the x-axis. Essentially, lower p-values, approaching 0, indicate instances where the forecast on the x-axis outperforms the forecast on the y-axis by a significant margin.}
    \label{fig:DM_Test}
\end{figure*}

Figure \ref{fig:DM_Test} stands out as a pivotal discovery in this study. As illustrated, the PSP method surpasses all other methods in performance. Applying the Diebold and Mariano (DM) test, employing the Mean Squared Error (MSE) as the loss function, affirms the superior accuracy of the PSP method in predictions compared to the other two methods, aligning with our expectations. Incorporating local volatilities instead of a single reference volatility proves to be a value-addition. Furthermore, the utilization of volatility, whether reference or local, demonstrates better results than not incorporating volatility at all, as seen in the Constant Volatility model. All drawn inferences hold significance, underscoring the validity of our findings.

\subsection{Performance-based Evaluation}\label{sec:Performance_based}

The Ranking Model proposed by \cite{csener2012ranking}, chosen for evaluating VaR calculation methods, introduces a penalty mechanism for each day where the forecasted VaR deviates from the realized return. The cumulative penalty across the entire period becomes instrumental in identifying the superior method. Notably, this model stands out for its unique approach—it eschews statistical tests and opts for scoring each VaR calculation method based on its performance. This characteristic allows the Ranking Model to concurrently compare several methods, setting it apart from previous models. Given our dataset's limitations for conducting robust statistical tests, the model's capacity to assess multiple methods simultaneously proves crucial for our analysis. 

Table \ref{tab:results_ranks} presents the results derived from the ranking model. Notably, the PSP method exhibits a relatively weaker performance for lower confidence intervals. This is primarily attributed to the method's higher frequency of violations. As previously discussed, the increased number of violations results in diminished accuracy from a regulatory standpoint but enhanced accuracy from a firm's perspective. In this scenario, the quantity and severity of violations lead to a lower rank for the PSP method. Conversely, the PSP method excels for higher confidence intervals, securing the top position. The fewer violations observed in the Const Vol and VIX models contribute to higher penalties, particularly from a firm's viewpoint. An intriguing observation is the relative stability of the Const Vol model, which consistently holds the second position across all confidence intervals.

\begin{table}
\begin{center}
\begin{tabular}{*5c}
\hline 
\hline
\toprule
Confidence&   Method   & Penalty & Percentage & Rank \\
\hline
90\%      & PSP        & $3.93 \times 10^{-2}$      &   38.53  & 3    \\
          & Const. Vol & $3.40 \times 10^{-2}$      &   33.33  & 2    \\
          & VIX        & $2.87 \times 10^{-2}$      &   28.14  & 1    \\
\hline
95\%      & PSP        & $1.46 \times 10^{-2}$      &   31.33  & 1    \\
          & Const. Vol & $1.54 \times 10^{-2}$      &   33.05  & 2    \\
          & VIX        & $1.66 \times 10^{-2}$      &   35.62  & 3    \\
\bottomrule
\hline
\hline
\end{tabular}
\caption{The performance of the PSP method in estimating VaR with 90\% and 95\% confidence levels is assessed through conditional coverage, unconditional coverage, and independence tests.}
\label{tab:results_ranks}
\end{center}
\end{table}

\section{Conclusion}\label{conclusion}

The computation of the value of non-linear instruments within derivative portfolios poses a challenge, demanding a delicate balance between precision and computational efficiency. In the initial section of this paper, we introduce an approach for PnL calculation that aims to enhance accuracy without incurring excessive computational costs. Subsequently, our performance evaluation encompasses unconditional coverage, independence, conditional coverage, the Diebold-Mariano test, and the Ranking Model. Const Vol and the VIX serve as benchmarks, with Const Vol representing a straightforward overall model and the VIX model simplifying the consideration of volatility.

In our proposed method, we initiate the process by extracting the volatility surface for each day from option prices. Subsequently, volatility shocks are computed based on historical data on the volatility surface, resulting in a database of projected volatility surfaces. Simultaneously, employing Monte Carlo simulation, we generate next-day prices for the underlying asset. These volatility shocks and next-day prices are then input into the Black-Scholes formula pricing model to obtain forecasts for option prices the following day.

Numerical tests reveal the superior performance of the PSP method, particularly at high confidence levels (e.g., 95\%). The DM test emphasizes the effectiveness of localized volatility models over reference volatility, capturing market conditions more accurately. Furthermore, incorporating volatility, regardless of its form (reference or localized), demonstrates potential improvements in forecasting accuracy. 

Lastly, the central component of the PSP method is the historical simulation block. We have incorporated Filtered Historical Simulation (FHS) to enhance accuracy by considering the relative importance of various scenarios through a weighting procedure in the HS block.


\appendix

\setcitestyle{numbers}
\bibliographystyle{cas-model2-names}
\bibliography{refs}

\begin{thebibliography}{31}
\expandafter\ifx\csname natexlab\endcsname\relax\def\natexlab#1{#1}\fi
\providecommand{\url}[1]{\texttt{#1}}
\providecommand{\href}[2]{#2}
\providecommand{\path}[1]{#1}
\providecommand{\DOIprefix}{doi:}
\providecommand{\ArXivprefix}{arXiv:}
\providecommand{\URLprefix}{URL: }
\providecommand{\Pubmedprefix}{pmid:}
\providecommand{\doi}[1]{\href{http://dx.doi.org/#1}{\path{#1}}}
\providecommand{\Pubmed}[1]{\href{pmid:#1}{\path{#1}}}
\providecommand{\bibinfo}[2]{#2}
\ifx\xfnm\relax \def\xfnm[#1]{\unskip,\space#1}\fi
\bibitem[{Arslan et~al.(2009)Arslan, Eid, El~Khoury and Roth}]{arslan2009gamma}
\bibinfo{author}{Arslan, M.}, \bibinfo{author}{Eid, G.},
  \bibinfo{author}{El~Khoury, J.}, \bibinfo{author}{Roth, J.},
  \bibinfo{year}{2009}.
\newblock \bibinfo{title}{The gamma vanna volga cost framework for constructing
  implied volatility curves}.
\newblock \bibinfo{journal}{Unpublished working paper. Deutsche Bank} .
\bibitem[{Aurell(2014)}]{aurell2014svi}
\bibinfo{author}{Aurell, A.}, \bibinfo{year}{2014}.
\newblock \bibinfo{title}{The svi implied volatility model and its
  calibration}.
\bibitem[{Black and Scholes(1973)}]{black1973pricing}
\bibinfo{author}{Black, F.}, \bibinfo{author}{Scholes, M.},
  \bibinfo{year}{1973}.
\newblock \bibinfo{title}{The pricing of options and corporate liabilities}.
\newblock \bibinfo{journal}{Journal of political economy} \bibinfo{volume}{81},
  \bibinfo{pages}{637--654}.
\bibitem[{Carr and Wu(2020)}]{carr2020option}
\bibinfo{author}{Carr, P.}, \bibinfo{author}{Wu, L.}, \bibinfo{year}{2020}.
\newblock \bibinfo{title}{Option profit and loss attribution and pricing: A new
  framework}.
\newblock \bibinfo{journal}{The Journal of Finance} \bibinfo{volume}{75},
  \bibinfo{pages}{2271--2316}.
\bibitem[{Chan(2017)}]{chan2017machine}
\bibinfo{author}{Chan, E.P.}, \bibinfo{year}{2017}.
\newblock \bibinfo{title}{Machine trading: Deploying computer algorithms to
  conquer the markets}.
\newblock \bibinfo{publisher}{John Wiley \& Sons}.
\bibitem[{Clark(2011)}]{clark2011foreign}
\bibinfo{author}{Clark, I.J.}, \bibinfo{year}{2011}.
\newblock \bibinfo{title}{Foreign exchange option pricing: a practitioner's
  guide}.
\newblock \bibinfo{publisher}{John Wiley \& Sons}.
\bibitem[{Daglish et~al.(2007)Daglish, Hull and Suo}]{daglish2007volatility}
\bibinfo{author}{Daglish, T.}, \bibinfo{author}{Hull, J.},
  \bibinfo{author}{Suo, W.}, \bibinfo{year}{2007}.
\newblock \bibinfo{title}{Volatility surfaces: theory, rules of thumb, and
  empirical evidence}.
\newblock \bibinfo{journal}{Quantitative Finance} \bibinfo{volume}{7},
  \bibinfo{pages}{507--524}.
\bibitem[{Dumas et~al.(1998)Dumas, Fleming and Whaley}]{dumas1998implied}
\bibinfo{author}{Dumas, B.}, \bibinfo{author}{Fleming, J.},
  \bibinfo{author}{Whaley, R.E.}, \bibinfo{year}{1998}.
\newblock \bibinfo{title}{Implied volatility functions: Empirical tests}.
\newblock \bibinfo{journal}{The Journal of Finance} \bibinfo{volume}{53},
  \bibinfo{pages}{2059--2106}.
\bibitem[{Dupire(1997)}]{dupire1997pricing}
\bibinfo{author}{Dupire, B.}, \bibinfo{year}{1997}.
\newblock \bibinfo{title}{Pricing and hedging with smiles}.
\newblock \bibinfo{journal}{Mathematics of derivative securities}
  \bibinfo{volume}{1}, \bibinfo{pages}{103--111}.
\bibitem[{Gershon(2018)}]{gershon2018model}
\bibinfo{author}{Gershon, D.}, \bibinfo{year}{2018}.
\newblock \bibinfo{title}{Model independent multi-asset volatility smile with
  empirical confirmation}.
\newblock \bibinfo{journal}{Available at SSRN 3214115} .
\bibitem[{Glasserman et~al.(2000)Glasserman, Heidelberger and
  Shahabuddin}]{glasserman2000efficient}
\bibinfo{author}{Glasserman, P.}, \bibinfo{author}{Heidelberger, P.},
  \bibinfo{author}{Shahabuddin, P.}, \bibinfo{year}{2000}.
\newblock \bibinfo{title}{Efficient monte carlo methods for value-at-risk} .
\bibitem[{Hendricks(1996)}]{hendricks1996evaluation}
\bibinfo{author}{Hendricks, D.}, \bibinfo{year}{1996}.
\newblock \bibinfo{title}{Evaluation of value-at-risk models using historical
  data}.
\newblock \bibinfo{journal}{Economic policy review} \bibinfo{volume}{2}.
\bibitem[{Homescu(2011)}]{homescu2011implied}
\bibinfo{author}{Homescu, C.}, \bibinfo{year}{2011}.
\newblock \bibinfo{title}{Implied volatility surface: Construction
  methodologies and characteristics}.
\newblock \bibinfo{journal}{arXiv preprint arXiv:1107.1834} .
\bibitem[{Hong et~al.(2014)Hong, Hu and Liu}]{hong2014monte}
\bibinfo{author}{Hong, L.J.}, \bibinfo{author}{Hu, Z.}, \bibinfo{author}{Liu,
  G.}, \bibinfo{year}{2014}.
\newblock \bibinfo{title}{Monte carlo methods for value-at-risk and conditional
  value-at-risk: a review}.
\newblock \bibinfo{journal}{ACM Transactions on Modeling and Computer
  Simulation (TOMACS)} \bibinfo{volume}{24}, \bibinfo{pages}{1--37}.
\bibitem[{Hull(2012)}]{hull2012risk}
\bibinfo{author}{Hull, J.}, \bibinfo{year}{2012}.
\newblock \bibinfo{title}{Risk management and financial institutions,+ Web
  Site}. volume \bibinfo{volume}{733}.
\newblock \bibinfo{publisher}{John Wiley \& Sons}.
\bibitem[{Hull and White(1987)}]{hull1987pricing}
\bibinfo{author}{Hull, J.}, \bibinfo{author}{White, A.}, \bibinfo{year}{1987}.
\newblock \bibinfo{title}{The pricing of options on assets with stochastic
  volatilities}.
\newblock \bibinfo{journal}{The journal of finance} \bibinfo{volume}{42},
  \bibinfo{pages}{281--300}.
\bibitem[{Hull(2003)}]{hull2003options}
\bibinfo{author}{Hull, J.C.}, \bibinfo{year}{2003}.
\newblock \bibinfo{title}{Options futures and other derivatives}.
\newblock \bibinfo{publisher}{Pearson Education India}.
\bibitem[{Jorion(1996)}]{jorion2007value}
\bibinfo{author}{Jorion, P.}, \bibinfo{year}{1996}.
\newblock \bibinfo{title}{Value at risk: the new benchmark for controlling
  derivatives risk}.
\newblock \bibinfo{publisher}{The McGraw-Hill Companies, Inc.}
\bibitem[{Lago et~al.(2021)Lago, Marcjasz, {De Schutter} and
  Weron}]{epftoolbox}
\bibinfo{author}{Lago, J.}, \bibinfo{author}{Marcjasz, G.},
  \bibinfo{author}{{De Schutter}, B.}, \bibinfo{author}{Weron, R.},
  \bibinfo{year}{2021}.
\newblock \bibinfo{title}{Forecasting day-ahead electricity prices: A review of
  state-of-the-art algorithms, best practices and an open-access benchmark}.
\newblock \bibinfo{journal}{Applied Energy} \bibinfo{volume}{293},
  \bibinfo{pages}{116983}.
\newblock \DOIprefix\doi{https://doi.org/10.1016/j.apenergy.2021.116983}.
\bibitem[{Mandelbrot and Taylor(1967)}]{mandelbrot1967distribution}
\bibinfo{author}{Mandelbrot, B.}, \bibinfo{author}{Taylor, H.M.},
  \bibinfo{year}{1967}.
\newblock \bibinfo{title}{On the distribution of stock price differences}.
\newblock \bibinfo{journal}{Operations research} \bibinfo{volume}{15},
  \bibinfo{pages}{1057--1062}.
\bibitem[{Mercurio(2007)}]{mercuriovanna}
\bibinfo{author}{Mercurio, A.C.F.}, \bibinfo{year}{2007}.
\newblock \bibinfo{title}{The vanna-volga method for implied volatilities:
  tractability and robustness} .
\bibitem[{Niu(2015)}]{niu2015no}
\bibinfo{author}{Niu, L.Q.}, \bibinfo{year}{2015}.
\newblock \bibinfo{title}{No arbitrage conditions and characters of implied
  volatility surface: A review for implied volatility modelers}.
\newblock \bibinfo{journal}{Available at SSRN 2672416} .
\bibitem[{Pritsker(2006)}]{pritsker2006hidden}
\bibinfo{author}{Pritsker, M.}, \bibinfo{year}{2006}.
\newblock \bibinfo{title}{The hidden dangers of historical simulation}.
\newblock \bibinfo{journal}{Journal of Banking \& Finance}
  \bibinfo{volume}{30}, \bibinfo{pages}{561--582}.
\bibitem[{Roccioletti(2015)}]{roccioletti2015backtesting}
\bibinfo{author}{Roccioletti, S.}, \bibinfo{year}{2015}.
\newblock \bibinfo{title}{Backtesting value at risk and expected shortfall}.
\newblock \bibinfo{publisher}{Springer}.
\bibitem[{Roper(2010)}]{roper2010arbitrage}
\bibinfo{author}{Roper, M.}, \bibinfo{year}{2010}.
\newblock \bibinfo{title}{Arbitrage free implied volatility surfaces}.
\newblock \bibinfo{journal}{preprint} .
\bibitem[{{\c{S}}ener et~al.(2012){\c{S}}ener, Baronyan and
  Meng{\"u}t{\"u}rk}]{csener2012ranking}
\bibinfo{author}{{\c{S}}ener, E.}, \bibinfo{author}{Baronyan, S.},
  \bibinfo{author}{Meng{\"u}t{\"u}rk, L.A.}, \bibinfo{year}{2012}.
\newblock \bibinfo{title}{Ranking the predictive performances of value-at-risk
  estimation methods}.
\newblock \bibinfo{journal}{International Journal of Forecasting}
  \bibinfo{volume}{28}, \bibinfo{pages}{849--873}.
\bibitem[{Skiadopoulos et~al.(2000)Skiadopoulos, Hodges and
  Clewlow}]{skiadopoulos2000dynamics}
\bibinfo{author}{Skiadopoulos, G.}, \bibinfo{author}{Hodges, S.},
  \bibinfo{author}{Clewlow, L.}, \bibinfo{year}{2000}.
\newblock \bibinfo{title}{The dynamics of the s\&p 500 implied volatility
  surface}.
\newblock \bibinfo{journal}{Review of derivatives research}
  \bibinfo{volume}{3}, \bibinfo{pages}{263--282}.
\bibitem[{Verma et~al.(2008)Verma, Ertosun, Wang, Ambruster and
  Giesecke}]{Verma2008ImpliedVM}
\bibinfo{author}{Verma, S.}, \bibinfo{author}{Ertosun, G.M.},
  \bibinfo{author}{Wang, W.}, \bibinfo{author}{Ambruster, B.},
  \bibinfo{author}{Giesecke, K.}, \bibinfo{year}{2008}.
\newblock \bibinfo{title}{Implied volatility modeling}.
\newblock \URLprefix \url{https://api.semanticscholar.org/CorpusID:427538}.
\bibitem[{Whaley(2009)}]{whaley2009understanding}
\bibinfo{author}{Whaley, R.E.}, \bibinfo{year}{2009}.
\newblock \bibinfo{title}{Understanding the vix}.
\newblock \bibinfo{journal}{The Journal of Portfolio Management}
  \bibinfo{volume}{35}, \bibinfo{pages}{98--105}.
\bibitem[{Wolberg and Alfy(2002)}]{wolberg2002energy}
\bibinfo{author}{Wolberg, G.}, \bibinfo{author}{Alfy, I.},
  \bibinfo{year}{2002}.
\newblock \bibinfo{title}{An energy-minimization framework for monotonic cubic
  spline interpolation}.
\newblock \bibinfo{journal}{Journal of Computational and Applied Mathematics}
  \bibinfo{volume}{143}, \bibinfo{pages}{145--188}.
\bibitem[{Zamani et~al.(2022)Zamani, Chaghazardi and
  Arian}]{zamani2022pathwise}
\bibinfo{author}{Zamani, S.}, \bibinfo{author}{Chaghazardi, A.},
  \bibinfo{author}{Arian, H.}, \bibinfo{year}{2022}.
\newblock \bibinfo{title}{Pathwise grid valuation of fixed-income portfolios
  with applications to risk management}.
\newblock \bibinfo{journal}{Heliyon} \bibinfo{volume}{8}.

\end{thebibliography}
\setcitestyle{numbers} 

\end{document}